\documentclass[12pt,a4paper,english,amssymb,nofootinbib,superscriptaddress]{revtex4-2}
\usepackage{lmodern}
\usepackage{lmodern}
\usepackage[T1]{fontenc}
\usepackage[latin9]{inputenc}
\usepackage{color}
\usepackage{babel}
\usepackage{verbatim}
\usepackage{amsmath}
\usepackage{amssymb}
\usepackage{esint}
\usepackage{graphicx}
\usepackage{subfigure}

\usepackage[unicode=true,pdfusetitle,
 bookmarks=true,bookmarksnumbered=false,bookmarksopen=false,
 breaklinks=false,pdfborder={0 0 1},backref=false,colorlinks=true]
 {hyperref}
\hypersetup{
 linkcolor=blue,citecolor=blue}

\makeatletter

\pdfpageheight\paperheight
\pdfpagewidth\paperwidth

\@ifundefined{textcolor}{}
{%
 \definecolor{BLACK}{gray}{0}
 \definecolor{WHITE}{gray}{1}
 \definecolor{RED}{rgb}{1,0,0}
 \definecolor{GREEN}{rgb}{0,1,0}
 \definecolor{BLUE}{rgb}{0,0,1}
 \definecolor{CYAN}{cmyk}{1,0,0,0}
 \definecolor{MAGENTA}{cmyk}{0,1,0,0}
 \definecolor{YELLOW}{cmyk}{0,0,1,0}
}

\usepackage{babel}
\usepackage{babel}
\usepackage{babel}
\usepackage{babel}
\usepackage{babel}
\usepackage{babel}
\@ifundefined{definecolor}{color}{}
\usepackage{babel}

\usepackage{latexsym}\usepackage{bm}

\makeatother

\begin{document}
\title{Fluid-membrane Descriptions of Various  Black Holes}

\author{Cagdas Ulus Agca}
\email{ulusagca@metu.edu.tr}

\author{Bayram Tekin}
\email{btekin@metu.edu.tr}

\affiliation{{\small{}Department of Physics,}\\
 {\small{}Middle East Technical University, 06800 Ankara, Turkey}}

\begin{abstract}
\noindent
The membrane paradigm of black holes is an effective theory that replaces the event horizon with a fictitious yet tangible fluid. It has provided us with valuable insights, especially in understanding the environment of black holes. The paradigm establishes a fluid/gravity correspondence that allows the computation of the thermal properties of the black hole in terms of the transport coefficients of the fluid. Recently, we showed that this is a van der Waals-type fluid for Kerr and especially for Johannsen-Psaltis black holes. Here, we use the paradigm to study the properties of various black holes in different dimensions to test the predictive capacity of effective theory. Among these, 
for the BTZ black holes, the paradigm gives a physical bulk viscosity, unlike the generic examples, for which the paradigm works with a negative bulk viscosity. For  Lorentz-violating black holes, we demonstrate that the parameter for Lorentz violation is seen as a hair under the paradigm, which shifts the ergoregion where the fluid pressure diverges. It might have a consequence for black hole jets. For asymptotically safe quantum-corrected black holes, the paradigm detects the final state of evaporation, i.e., a remnant with a correct value for the mass that still has an event horizon. Finally, we check the paradigm with stringy black holes that have dilatonic and axionic charges, and the fluid produces the known results. 
\end{abstract}
\maketitle

\tableofcontents
\section{Introduction}
Our recent work \cite{Ulus}, in trying to develop a membrane paradigm approach to the Johannsen-Psaltis black hole \cite{Johannsen:2011dh}, a parametrically deviated phenomenologically motivated version of the Kerr black hole, gave some interesting results, such as the role of the ergoregion in the formation of black hole jets. The effective fluid that mimics the black hole, when we analytically continue into the whole space, including the interior of the black hole, shows a generalized van der Waals fluid-type behavior. These observations prompted us to develop the membrane paradigm of various black holes in different dimensions and to understand how ubiquitous the van der Waals fluid is, hence this work. 

For an external observer, the spacetime inside the event horizon of the black hole, including the event horizon, is out of the reach of observations. This basic fact led researchers to introduce an effective, tangible description of the black hole by a timelike fluid membrane infinitesimally close to the event horizon. Being a timelike codimension one hypersurface, the fluid has observable properties local in time, unlike the event horizon, which is highly nonlocal in time \cite{Finkelstienoriginal}. Of course, in the limit of going from a timelike surface to a null one, a proper regularization of certain functions, such as the lapse function in the metric, is needed. That is part of the paradigm's limiting process. The origins of the membrane paradigm\footnote{The name "membrane paradigm" was coined in  \cite{membrane_THORNE}, and the authors worked out many details about the membrane description.} approach can be traced back to 1979 when  Damour published \cite{Damour_Actio_membranes}. The membrane has electrical conductivity, shear $\&$ bulk viscosities, and other transport coefficients \cite{membrane_THORNE}. A careful calculation on the horizon led to Ohm's law, Joule's law, and the non-relativistic Navier-Stokes equation \cite{MembraneHorizonsBlackHolesNewClothes}. Recently, an action formulation of the membrane paradigm was given by Parikh and Wilczek \cite{Parikh_original}, who considered the Gibbons-Hawking-York (GHY) boundary term, not as a counterterm to be added to the boundary at infinity, but as a boundary term on a timelike hypersurface that envelopes the event horizon. Since the surface defining the membrane is not null, it has a nondegenerate metric, which approximates the behavior of the true event horizon. The membrane approach has been carried out for four-dimensional black holes. More recently, \cite{Silvestrini:2025lbe} shows that some ideas in the membrane paradigm can be extended to non-singular compact objects. Tidal Love numbers and quasi-normal modes can be extracted solely by the paradigm's tools. This also provided a future outlook, suggesting that one might also examine the paradigm of topological stars, i.e., geometrically transitioned black hole solutions to smooth compact objects \cite{Bah:2020ogh}. Here we shall find the membrane that mimics the Banados-Teitelboim-Zanelli (BTZ) black hole \cite{Banados:1992wn}: a $(2+1)$ dimensional Einstein spacetime, with a negative cosmological constant, that admits a black hole structure without bulk degrees of freedom; moreover acts as a final state of a collapsing matter \cite{Carlip:1995qv}. The existence of the cosmological constant drastically affects the construction of the membrane. For example, one requires a regularized energy-momentum tensor on the boundary of the Anti-de Sitter spacetime. Furthermore, we shall explore various black holes, such as  Lorentz-violating, asymptotically safe, and string-inspired black holes, within the context of the membrane paradigm to see how well this effective theory captures the physics of these beyond Einstein's gravity where we found its success in our previous work \cite{Ulus} with the parametrically deviated black hole solutions.

 The layout of this paper is as follows. In section (\ref{BTZ1}), and section (\ref{BTZ2}), we reformulate the $2+1+1$ splitting ideas in $4d$ black holes to the $1+1+1$ splitting of $(2+1)$-dimensional BTZ black holes and identify their transport coefficients as if the stretched horizon is a viscous Newtonian fluid. Moreover, we also apply the paradigm to the rotating version of the black hole. The rotating BTZ black hole also has a zero bulk viscosity \cite{Arslaniev_original}. In section \eqref{LorentzViolation}, we extend the membrane paradigm's machinery to some Lorentz-violating spacetimes and show that the value of ergoregion radius changes with the Lorentz violation parameter and the phase transition of the fluid can be detected by the pressure value. This shows that the degree of freedom spontaneously broken by the Lorentz violating parameter $\ell$  radically affects all fluid coefficients, and the well-known values could be restored by setting $\ell=0$,  which is the no-violation limit. In Section \eqref{asymptoticalsafetyschwarzschild}, we explore the effects of running Newton's constant on the stretched horizon in asymptotically safe gravity. In the last section \eqref{stringyblackholes}, we consider string theory-inspired black holes in the paradigm and show that it reproduces the well-known results even in the boosted frame while accommodating both the axionic and the dilatonic degrees of freedom. In Appendix \eqref{appendix}, we give a brief review of the Newman-Janis (NJ) algorithm, which makes it easy to find a rotating counterpart of a given static black hole through analytical continuation. For the membrane paradigm to work, it requires a constraint on the value of the bulk viscosity, where both static and rotating bulk viscosities must be equal to each other. The applicability of the NJ algorithm for a static black hole implies a rotating cousin for it, where one can find the static bulk viscosity and employ the constraint on the rotating counterpart. However, the paradigm's transport coefficients do not give the correct values when the NJ algorithm is applied to them. Hence, static transport coefficients cannot generally imply the rotating ones through the algorithm. One needs to use the NJ algorithm to find the rotating black hole metric and then apply the paradigm's machinery. The NJ algorithm cannot translate the static fluid degrees of freedom to rotating ones. Moreover, we relegate the computational details to the appendices as some of the results are cumbersome (\ref{transportcoefificients}).

\section{Membrane  describing the static BTZ black hole}\label{BTZ1}

In what follows, we will construct gravitational membranes that effectively reproduce the observable properties of non-rotating and rotating BTZ, Lorentz-violating, asymptotically safe, and more generally charged black holes. The membrane paradigm has many ingredients that must be adequately defined before construction. In our recent paper \cite{Ulus}, we have given a fairly involved description of this, and hence, here, for ease of reading, we will only briefly recapitulate the necessary geometric quantities and the corresponding properties of the fluid membrane. See also \cite{Arslanaliev:2023vev,agca_2023} for further details. In no way do we claim novelty about the basic construction of the membrane. We are using the ideas developed over the years; especially, we shall employ the action formulation of the membrane by Parikh and Wilczek \cite{Parikh_original,MembraneHorizonsBlackHolesNewClothes}, which is best suited to pinpoint the transport coefficients.

Let $(\mathcal{M},g)$ be the $(2+1)$-dimensional spacetime that contains a black hole, by which we mean that there exists a codimension one null hypersurface $\mathcal{H}$ with a degenerate metric on it; and, for our case, with a spatial cross-section having the topology of $S^1$. The crux of the membrane paradigm is to posit the existence of a stretched horizon $\mathcal{H}_s$ that is timelike with a non-degenerate metric $h$. The dynamical behavior of the stretched horizon located arbitrarily close to the true horizon $\mathcal{H}$ is that of a viscous fluid, such that in the limit to the true horizon, it correctly gives the thermodynamical quantities of the black hole. The stretched horizon is observable, unlike the true horizon, which is infinitely nonlocal in time and is beyond the scope of observations for transient observers like us.

Unlike the construction in asymptotically flat spacetimes, the BTZ black hole exists only in anti-de Sitter spacetimes. It requires an additional term in the quasi-local stress tensor on the stretched horizon. To reproduce the known results via the membrane paradigm, one should perform a $(1+1+1)$-decomposition of the black hole metric. The immense acceleration, in the limit to the true horizon, makes some of the extensive thermodynamic quantities divergent. Therefore, as in the four-dimensional construction, one needs some way of regularization, best done with the help of the lapse function $N(r)$, which induces a cutoff to the surface gravity such that it reproduces finite results. This regularization is needed, as one cannot smoothly get a null surface from a time-like surface. 

The 2+1-dimensional Banados-Teitelboim-Zanelli (BTZ) black hole has proved to be a remarkably efficient tool to test ideas about quantum aspects of gravity and holography. It is a simpler setting of gravity since there are no bulk-propagating degrees of freedom. Here, we provide a classical fluid-membrane description of this geometry, both in static and rotating versions, that defines a dynamical viscous Newtonian fluid with nontrivial transport coefficients, mimicking the black hole. The membrane paradigm for black holes in four or more dimensions usually yields fluids with a negative bulk viscosity, which is a problem because it yields decreasing total entropy in thermal processes according to Stokes' hypothesis. However, the (rotating) BTZ black hole has a zero bulk viscosity and is amenable to a description of a physically viable fluid membrane in thermal equilibrium.

We start with the {\it static} BTZ black hole in the usual $(t,r,\phi)$ coordinates
\begin{equation}
    ds^2=-f(r)dt^2+ \frac{dr^2}{f(r)}+r^2d\phi^2, \hskip 1 cm f(r)=-m+\frac{r^2}{\ell^2},
    \label{metric1}
\end{equation}
which is locally $AdS_3$ with the cosmological constant $\Lambda=-\frac{1}{\ell^2}$  \cite{Banados:1992wn}. The membrane paradigm requires an event horizon with a compact cross-section, which for this case is $S^1$ and is located at the largest root of $g^{rr}=0$. The metric (\ref{metric1}) can be rewritten as 
    \begin{gather}
        ds^2=-u_\mu u_\nu dx^\mu dx^\nu+n_\mu n_\nu dx^\mu dx^\nu+\gamma_\mu\gamma_\nu dx^\mu dx^\nu,
    \end{gather}
where the $\gamma_\mu$ part denotes the $1d$ space-like cross section of the stretched horizon $\mathcal{H}_s$, and one has 
\begin{align}
    u_\mu dx^\mu=\sqrt{f}dt,&&
    n_\mu dx^\mu=\frac{1}{\sqrt{f}}dr,&&
    \gamma_{\mu}dx^\mu =r d\phi,&&
\end{align}
and the induced metric on ${\cal{H}}_s$ is
\begin{equation}
    ds^2_{\mathcal{H}_s}=-fdt^2+r^2d\phi^2.
\end{equation}
The extrinsic curvature (and its trace) of the stretched horizon follows from the definition  $K_{\mu\nu}:=  \nabla_\mu n_\nu$ and reads, respectively, as 
\begin{equation}
  K_{\mu\nu}= -\sqrt{f}\begin{pmatrix}
\frac{1}{2}\partial_r f & 0 & 0 \\
0 & 0 & 0\\
  0& 0  & -r 
\end{pmatrix}, \hskip 1 cm 
    K=\frac{\sqrt{f}}{r}+\frac{1}{2}\frac{\partial_r f}{\sqrt{f}}.
\end{equation}
These expressions will be used in the computation of the {\it boundary stress tensor}, which is the main object of interest, and we do this next.
\subsection{ Stress tensor using the boundary action}

The method proposed in \cite{Parikh_original} assumes a boundary at a finite distance such that the extremal action principle works with the added surface term: 
\begin{equation}
    \delta_g S_{\text{total}}=\delta_g \,S_{{\text{total,1}}}+ \delta_g\,S_{\text{total,2}},\notag
\end{equation}
\begin{equation}
    \delta_g S_{\text{total}}=\delta_g (S_{{\text{in}}}+ S_{{\text{surface}}})+\delta_g(S_{\text{out}}-S_{\text{surface}}),
\end{equation}
where the surface term refers to the black hole boundary, which is now represented effectively by the membrane, the first and the second parts of the action variations are assumed to vanish separately \cite{MembraneHorizonsBlackHolesNewClothes}. 

For the BTZ black hole, the spacetime has a non-zero cosmological constant. From the perspective of \cite{Parikh_original}, the additional cosmological constant term can be viewed as a zeroth-order correction that is canceled by an extra counter-term. However, one can first construct the stress tensor, then impose the junction conditions on the boundary \cite{Israel_junction_condition}, and then add the counterterm. 
For this purpose, let us define a $d+1$-dimensional gravity without a cosmological constant and a non-vanishing boundary contribution, i.e., the Gibbons-Hawking boundary term (in the units  $G_N=1$, $c=1$)
\begin{equation}
    S_{\text{total,1}}=\frac{1}{16\pi}\int_\mathcal{M} d^{d+1}x\sqrt{-g} \,R+\frac{1}{8\pi}\oint_\mathcal{\partial M} d^{d}x\sqrt{\pm h}\,K + S_{\text{surface}},
\end{equation}
\begin{equation}
    \delta S_{\text{in}}=\frac{1}{16\pi} \int_{\partial\mathcal{ M}} d^dx\sqrt{-h}\big(Kh_{\mu\nu}-K_{\mu\nu}\big)\delta g^{\mu\nu}.
    \label{stretchedhorizontensor}
\end{equation}
One finds the variation of the action on-shell to be the Brown-York quasi-local stress tensor. Further corrections on that quasi-local stress tensor require higher derivative corrections on the gravity action. Then, one has $t^{\text{stretched}}_{\mu\nu}=\frac{1}{8\pi}\left(Kh_{\mu\nu}-K_{\mu\nu}\right)$. On the stretched horizon $g^{\mu\nu}\vert_{\mathcal{H}_s}=h^{\mu\nu}$. To cancel the above non-zero boundary variation term, one must add the following:
\begin{equation}
    \delta S_{\text{surface}}=-\frac{1}{2}\int d^d x\sqrt{-h}\,t^{\text{stretched}}_{\mu\nu}\delta h^{\mu\nu}.
\end{equation}
We need the Israel junction condition to resolve the discontinuity on the given boundary. \footnote{One can find an analogous example in electrodynamics: Since the surface charge induces a discontinuity in the observables of the theory on a surface, resolving that discontinuity requires imposing boundary conditions on the surface. In our case, $t^{\text{stretched}}_{\mu\nu}$ induces a discontinuity in the stretched horizon's $K_{\mu\nu}$, i.e  the momentum conjugate of $h_{\mu\nu}$ \cite{Parikh_original}, which can be solved by the Israel junction condition  \cite{Israel_junction_condition}.}

\begin{equation}
    t^{\text{stretched}}_{\mu\nu}=\frac{1}{8\pi}\left([K]h_{\mu\nu}-[K]_{\mu\nu}\right),
\end{equation}
where $[K]=K^+-K^-$ such that $[K]$ is the difference between the external universe embedding $\mathcal{H}_s$. We should identify $K^-=0$ so that the stretched horizon interior to the black hole side is a flat embedding. This flat embedding is suggested by the paradigm that assumes there is no \textit{phenomenological} reason to worry about the inside of a black hole since one cannot send information to the outside observers \cite{MembraneHorizonsBlackHolesNewClothes} beyond the horizon. Hence, assuming the interior of a black hole as a flat or curved manifold does not change what the outside observer sees.
 One can immediately see that $t_{\text{stretched}}^{\mu\nu}$ is not covariantly conserved because of the non-trivial source term; instead, one has
\begin{equation}
    D_\nu t_{\text{stretched}}^{\mu\nu}=-{h}{^\mu}_{\lambda}T^{\lambda\gamma}n_\gamma.
\end{equation}
This expression is none other than the Damour-Navier-Stokes equation, which allows one to see the gravitational membrane as a fluid-like surface. \cite{MembraneView}. We can further decompose ${K}{^\mu}_{\nu}$ in terms of the observables of the gravitational theory, i.e., surface gravity $\kappa$, null expansion $\Theta$. The shear tensor $\sigma_{AB}$ comes from the extrinsic curvature ${k}{^A}_{B}$ of the space-like cross section of $\mathcal{H}_s$. To this end, let $k_{AB}$ be the extrinsic curvature of the codimension-2 space-like section of $\mathcal{H}_s$, with $A, B$ running over the indices of the codimension-2 surface. Then, one has 
\begin{equation}
    K_{AB}=\sigma_{AB}+\frac{1}{2}\gamma_{AB}\Theta,
    \label{2.132}
\end{equation}
leading to the stretched horizon stress tensor \cite{MembraneHorizonsBlackHolesNewClothes}:
\begin{flalign}
t^{AB}_{\text{stretched}}&=\frac{1}{8\pi}\left(-\sigma^{AB}+\gamma^{AB}(\frac{1}{2}\Theta+\kappa)\right).
\end{flalign}
Based on this construction, we will modify the stretched horizon stress tensor such that it also incorporates the correct regularization of both the boundary at infinity and the membrane at a finite distance for asymptotically $AdS_3$ spacetimes, especially in the context of the BTZ black holes.

\subsection{Boundary stress tensor for the static BTZ black hole}

As is well known, a local stress-energy tensor does not exist for the gravitational field because of the equivalence principle. Therefore, one defines a "quasi-local" tensor for a finite region of the spacetime. By definition, this tensor can be written as $t^{\mu\nu}=-\frac{2}{\sqrt{-h}}\frac{\delta S}{\delta h_{\mu\nu}}$, where $h_{\mu\nu}$ is the metric on ${\cal{H}}_s$.

We constructed the necessary tensors for defining the quasi-local stress tensor above, enabling us to determine the quasi-local stress tensor without divergence at infinity. Nevertheless, before that, let us revisit the construction of the transport coefficients in a known example, that is, the static Schwarzschild metric in $3+1$ dimensions:
\begin{equation}
    ds^2=-fdt^2+f^{-1}dr^2+r^2d\Omega_2, \hskip 1 cm f=1-\frac{2 m}{r}.
\end{equation}
In the $1+1+2$-splitting\footnote{This construction follows \cite{Arslaniev_original,Arslanaliev:2023vev}.}, the metric reads as 
\begin{equation}
    ds^2=-u_\mu u_\nu dx^\mu dx^\nu+n_\mu n_\nu dx^\mu dx^\nu+\gamma_{\mu\nu}dx^\mu dx^\nu,
\end{equation}
 where $u_\mu$ and $n_\mu$  satisfy $u_\mu u^\mu=-1$,  $n_\mu n^\mu=1$ while  $u_\mu n^\mu=0$ on the stretched horizon $\mathcal{H}_s$. At the event horizon $\mathcal{H}_{r\rightarrow 2 m}$, the vectors $u$ and $n$ should be null. On the $2d$ surface, we will use the coordinates $\{A,B\}=\{\theta,\phi\}$ hence $\gamma_{AB}=\text{diag}(r^2,r^2\sin^2\theta)$. 
The extrinsic curvature tensor for this geometry reads as 
\begin{equation}
K_{\mu\nu}=-\frac{1}{2}\frac{\partial_r f}{\sqrt{f}} u_\mu u_\nu+\frac{\sqrt{f}}{r}\gamma_{\mu\nu},
\end{equation}
of which the trace is $K=\frac{1}{2}\frac{\partial_r f}{\sqrt{f}}+\frac{2\sqrt{f}}{r}$.
The extrinsic curvature of $\mathcal{H}_s$ can be identified by choosing the lapse function $N=\sqrt{f}$ as a renormalization factor:
\begin{equation}
    K_{\mu\nu}\longrightarrow N^{-1}(k_{\mu\nu}+\kappa u_\mu u_\nu),
    \label{extrinsiccurvature}
\end{equation}
where $k_{\mu\nu}=\gamma_{\mu A}\gamma_{\nu B}k^{AB}$ is the extrinsic curvature of the $2d$ surface; and $\kappa$ is the surface gravity \cite{Arslaniev_original}. In the limit $N\rightarrow  0$, the extrinsic curvature of the stretched horizon converges to that of the event horizon, and $K_{\mu\nu}$ becomes proportional to the surface gravity. The trace of $K_{\mu\nu}$ diverges since $f$ vanishes at $r=2m$:
\begin{flalign}
    &\lim_{N\to 0} K=\left.\frac{1}{2}\frac{\partial_r f}{\sqrt{f}} 
  \right\vert_{2 m}\rightarrow
    \left.\text{Tr}(N^{-1}k_{\mu\nu}-N^{-1}\kappa u_\mu u_\nu)\right\vert_{2 m},\label{limitscalarcurvature}\\
    &\lim_{N\to 0} K_{tt}=-\left.\frac{1}{2}\frac{\partial_r f}{\sqrt{f}} \right\vert_{2 m}\rightarrow \left.N^{-1}\kappa\right\vert_{2 m}.\label{limitKtt}
\end{flalign}
The equations (\ref{extrinsiccurvature}),(\ref{limitscalarcurvature}),(\ref{limitKtt}) can be combined to find the stress tensor:
\begin{align}
    t^{\text{stretched}}_{\mu\nu}
    =\frac{1}{8\pi N}\left((\frac{1}{2}\Theta+\kappa)\gamma_{\mu\nu}-\Theta u_\mu u_\nu-\sigma_{\mu\nu}\right),
    \label{stretchedhorizonstresstensor}
\end{align}
which is to be compared with the energy-momentum tensor of a viscous fluid given as 
\begin{align}
    t^{\text{viscous}}_{\mu\nu}&=N^{-1}\rho u_\mu u_\nu+N^{-1}\gamma_{\mu A}\gamma_{\nu B}\big(P\gamma^{AB}-2\eta\sigma^{AB}-\zeta\Theta\gamma^{AB}\big)\label{viscoustensor}&\\&+\pi^A(\gamma_{\mu A}u_\nu+\gamma_{\nu B}u_\nu)\notag.&
    \label{viscousstresstensortensor}
\end{align}
Here $\rho$ is the energy density, $P$ is the pressure, $\Theta$ is the null geodesic expansion coefficient, $\zeta$ is the bulk viscosity, $\eta$ is the shear viscosity, $\pi^A$ is the momentum density, $\sigma^{AB}$ is the shear tensor of the fluid. If we  identify \eqref{stretchedhorizonstresstensor}  with \eqref{viscoustensor}, we get the following transport coefficients of the fluid:
\begin{equation}
    \rho=-\frac{1}{8\pi}\Theta ,  \quad\eta=\frac{1}{16\pi},
    \quad P=\frac{\kappa}{8\pi} ,\quad \zeta=-\frac{1}{16\pi},\quad
    \pi^A=0.
    \label{transportcoefficientsstatic}
\end{equation}
Since we also have 
\begin{equation}
    t^{\text{stretched}}_{\mu\nu}=\frac{1}{8\pi}\left(\left(\frac{1}{2}\frac{\partial_r f}{\sqrt{f}}+\frac{\sqrt{f}}{r}\right)\gamma_{\mu\nu}-\frac{2\sqrt{f}}{r}u_\mu u_\nu\right),
\end{equation}
by comparison, we get 
\begin{align}
\label{viscousstatic}
    \Theta=\frac{2}{r}f ,\quad\quad\quad \sigma_{AB}=0,\quad\quad\quad
    \kappa=\frac{\partial_r f}{2}&.
\end{align}
In particular, for the Schwarzschild geometry with $f=1-\frac{2 m}{r}$, we have:
\begin{equation}
    \left.\Theta\right\vert_{2 m}=0 ,\quad \left.\sigma_{AB}\right\vert_{2 m}=0,\quad
     \left.\kappa\right\vert_{2 m}=\frac{1}{4 m}.
\end{equation}

 Observe that the surface gravity $\kappa$, energy density $\rho$, pressure $P$, the null expansion $\Theta$, and the shear tensor $\sigma^{AB}$ will change when $f$ changes in different coordinates. However, $\eta$ and $\zeta$  are universal for spherical horizons; in particular, the value of the bulk viscosity is negative, showing that we are dealing with an unstable fluid.

The above example was for an asymptotically flat geometry. For the BTZ case,  we need to be careful and consider the action formulation of the theory with proper counterterms to make the on-shell action finite. 
 
Let us apply the procedure for the asymptotically $AdS$ spacetime in $2+1$ dimensions.
Given the action,
\begin{equation}
    S=\frac{1}{16\pi}\int_\mathcal{ M} d^{3}x\sqrt{-g}\, \left(R-\frac{2}{\ell^2}\right)+\frac{1}{8\pi}\int_\mathcal{\partial M} d^2 x\sqrt{-h}\, K-\frac{1}{8\pi}S_{\text{ct}}(h_{\mu\nu}),
\end{equation}
one can do the on-shell variation
\begin{equation}
    \delta S=\frac{1}{8\pi}\int_\mathcal{\partial M} d^2x\sqrt{-h}\,\, t_{\mu\nu}\,\,\delta h^{\mu\nu}-\frac{1}{8\pi}\int_\mathcal{\partial M} d^2x\,\,\frac{\delta L_{\text{ct}}}{\delta h^{\mu\nu}}\,\,\delta h^{\mu\nu},
\end{equation}
where the Gibbons-Hawking-York quasi-local tensor, i.e., the conjugate momentum of the gravitational field, is $t_{\mu\nu}=\frac{1}{16\pi}\left(K_{\mu\nu}-K h_{\mu\nu}\right)$. In this form, one can see that the boundary term diverges as $r\rightarrow\infty$. However, if one chooses the counterterm as $L_{\text{ct}}=\sqrt{-h}\frac{1}{\ell}$, in $AdS_3$, one gets a quasi-local tensor vanishing at infinity. So the regularized quasi-local stress tensor for the asymptotically $AdS_3$ becomes \cite{Balasubramanian:1999re}
\begin{equation}
    t_{\mu\nu}=\frac{1}{16\pi}\left(K_{\mu\nu}-K h_{\mu\nu}-\frac{1}{\ell}h_{\mu\nu}\right).
\end{equation}
Let us rewrite the stretched horizon stress tensor:
\begin{equation}
t^{\text{stretched}}_{\mu\nu}=\frac{1}{8 \pi\sqrt{f}}\left(\left(\frac{\partial_r f}{2}-\frac{\sqrt{f}}{\ell}\right)\gamma_{\mu}\gamma_{\nu}-\left(\frac{f}{r}-\frac{\sqrt{f}}{\ell}\right)u_\mu u_\nu\right),
\label{staticbtzstretchedhorizon}
\end{equation}
and from the geometric decomposition description, one can check that 
\begin{equation}
    t_{\mu\nu}=\frac{1}{8\pi N}\left(-\Theta u_\mu u_\nu+\kappa\gamma_\mu \gamma_\nu\right),
    \label{geometricdecompositonstaticbtz}
\end{equation}
where $N=\sqrt{f}$. Unlike the case for a $4d$ Schwarzschild black hole, the shear tensor is not even defined for a $1d$ space-like cross-section. Upon identification of \eqref{staticbtzstretchedhorizon} with \eqref{geometricdecompositonstaticbtz}, one arrives at
\begin{align}
    \Theta=-\frac{f}{r}+\frac{\sqrt{f}}{\ell},&&\kappa=\frac{\partial_r f}{2}-\frac{\sqrt{f}}{\ell},
\end{align}
which behave at the event horizon as 
\begin{align}
    \Theta=-\frac{-m+\frac{r_\text{H}^2}{\ell^2}}{r_\text{H}}+\frac{\sqrt{-m+\frac{r_\text{H}^2}{\ell^2}}}{\ell}=0,&&\kappa= \frac{r_\text{H}}{\ell^2}-\frac{\sqrt{-m+\frac{r_\text{H}^2}{\ell^2}}}{\ell}=\frac{\sqrt{ m}}{\ell},
    \label{nullexpansionsurfacegravityatthehorizon}
\end{align}
which match the known results. As one can see, when $r\rightarrow \infty$, the $r^2$ term in $f$ dominates, and both the null-expansion and the surface gravity vanish, which leads to a zero stress horizon stretched tensor at the asymptotic infinity, as derived. We might not include the extra correction to the boundary since the horizon radius is at a finite distance, but not at the asymptotic infinity. However, we will see that the corrections of the type $\frac{\sqrt{\Delta}}{8 \pi r \ell}$ will also regularize the behavior at infinity for the rotating solution. This shows that the membrane paradigm's regularization function on the stretched horizon regularizes both the null and asymptotically  $AdS$ boundary at once.

The static spacetime requires a vanishing momentum on the dual description; moreover, the shear vector is non-existent as expected in a $(2+1)$-dimensional spacetime. Mainly, transport coefficients depend on the underlying theory; however, the shear viscosity in Einstein's gravity is universal. Also, according to Stokes' Hypothesis \cite{stokeshypothesis}, black hole thermal states must be in thermal equilibrium if and only if $\zeta=0$, which simplifies the stretched horizon tensor. Here let us consider a $(d+1)$-dimensional spacetime with a $(d-1)$-spacelike cross-section, then the bulk viscosity reads as  $\zeta=-\frac{d-2}{8\pi(d-1)}$.
For a $3d$ spacetime, horizon dim$(\mathcal{H})=2d$ with a $1d$ cross-section. Hence $\zeta=0$ for $3d$ and 
\begin{align}
\pi^\mu=0,\quad\quad\zeta=0,\quad\quad
\sigma_{\mu}=0.
\end{align}
Therefore, the stress tensor of the viscous fluid reduces to a simplified form:
\begin{equation}
    t^{\text{viscous}}_{\mu\nu}=N^{-1}\rho u_\mu u_\nu+N^{-1}\gamma_\mu \gamma_A \gamma_\nu \gamma_B\left(P\gamma^A\gamma^B\right),
\end{equation}
where we can identify the non-vanishing coefficients as:
\begin{align}
    \rho=-\frac{1}{8\pi}\left(-\frac{f}{r}+\frac{\sqrt{f}}{\ell}\right),&&P=\frac{1}{8\pi}\left(\frac{\partial_r f}{2}-\frac{\sqrt{f}}{\ell}\right),&& \eta=\text{arbitrary}.
\end{align}
Since the shear tensor is non-existent, its coefficient is kept arbitrary for now. However, the membrane paradigm for $2+1$-dimensional geometries cannot fix the value of shear viscosity. Analyzing \eqref{viscoustensor}, one observes that the shear tensor of the fluid stress vanishes for both rotating and static configurations, and vanishing bulk viscosity allows one to find the energy density and pressure uniquely. However, it leaves the shear viscosity arbitrary. Since the universality of shear viscosity and entropy density ratio, i.e. $\frac{\eta}{s}=\frac{1}{4\pi}$ holds for Einstein space black holes without higher derivative corrections, as in the case of the BTZ black hole, one can fix the value for the shear tensor as $\eta=\frac{1}{16 \pi}$.
 As one can see, the coefficient in front of the null expansion depends solely on the dimension of the spacetime, i.e., $\frac{(d-2)f}{r}$. These are the correct values for the geometry at the event horizon found in the literature. The dual-fluid description of null expansion corresponds to the energy flux and the $00$ component of the stress-energy tensor. If one picks a more general metric function, the membrane paradigm can see the additional hair as long as the compact null-horizon condition is satisfied as a stable black hole. For example, if one considers the {\it linearly  charged} BTZ black hole \cite{Prasia:2016esx}
\begin{equation}
    f(r)=-m +\frac{r^2}{\ell^2} - q^2\ln{r^2},
\end{equation}
where the seed $f(r)$ is the chosen lapse function with two real roots. The Cauchy horizon is at  $r_\text{C}$, and the event horizon is at  $r_\text{H}$, given that $m\geq 1$. At $r_\text{H}$, the maximum value of the charge can be set as $q_{\text{max}}=\sqrt{2 r_\text{H}}$. It can be seen that for certain values of $q$, $m<0$, the horizons are intact. The transport coefficients of the linearly charged BTZ black holes are \cite{Carlip:1995qv}:
\begin{align}
    \Theta=\left.\frac{1}{r}\left(-m +\frac{r^2}{\ell^2} - q^2\ln{r^2}\right)\right\vert_{r\mapsto r_\text{H}}=0,&&\kappa=\left.\left(\frac{r}{\ell^2}-\frac{q^2}{r}\right)\right \vert_{r\mapsto r_\text{H}}.
    \label{nullexpansionsurfacegravitychargedbtz}
\end{align}
One can see that the surface gravity analysis coming from $\zeta=0$ agrees with those known in the literature \cite{surfacegravityBTZ}.

\section{Membrane description of the rotating BTZ black hole} \label{BTZ2}

The rotating BTZ black hole is again locally $AdS_3$, but globally, its causal structure is such that there is an event horizon and the spacetime is endowed with a mass $m$ and 
angular momentum $J$. The explicit form of the metric can be obtained in many ways, such as applying the Newman-Janis algorithm  \cite{Erbin:2016lzq} to a static BTZ solution \cite{Kim:1998iw}. For our purposes, the BTZ black hole is best described in the Boyer-Lindquist coordinates \cite{Carlip:1995qv}
\begin{equation}
    ds^2=-f dt^2-2a\left(1-f\right)dt d\phi+\frac{r^2}{\Delta}dr^2+\left(\left(r^2+a^2\right)+a^2\left(1-f\right)\right)d\phi^2.
    \label{blcoordinatedbtzblackhole}
\end{equation}
In the notation of \cite{Arslaniev_original}, the seeds can be decomposed as:
\begin{align}
    F_t^2=f,& & F^2_r=\frac{r^2}{\Delta},&&
    F_\phi^2=a^2 \left(1-f\right)+\left(a^2+r^2\right), && \omega=a \left(1-f\right) F^{-1}_t,
\end{align}
where $f=-m+\frac{r^2}{\ell^2}$, $\Delta=a^2+\frac{r^4}{\ell^2}-m r^2$ while cosmological constant is $\Lambda=-\frac{1}{\ell^2}$. This form of the BTZ black hole can also be seen as the section of the $4d$ Kerr black hole for $\theta=\frac{\pi}{2}$. This geometric equivalence will also show up in the transport coefficients of the corresponding fluid membranes. The metric given in \eqref{blcoordinatedbtzblackhole} is an Einstein space: i.e. $R_{\mu\nu}=2 \ell^{-2}g_{\mu\nu}$. One can make a coordinate transformation  $\Tilde{t}=t-a\phi$,  with $a=\frac{J}{2}$ \cite{Kim:1998iw} to arrive at the usual form of the BTZ metric. 
The outer horizon and the inner horizon are located at 
the roots of $\Delta=0$:
\begin{equation}
    r_{\text{H}_{\pm}}=\frac{\ell}{\sqrt{2}}\sqrt{ m\pm  \sqrt{m^2-\frac{4 a^2}{\ell^2}}}, \hskip 1 cm   | \ell m| \ge 2 a .
\end{equation}
As expected, the ergosphere coincides with the $a\rightarrow 0$ limit of the positive horizon radius namely, $r^\text{erg}_{\text{H}_+}\vert_{a\rightarrow 0}=\ell \sqrt{m}$. Note that the BTZ black hole in the Boyer-Lindquist coordinates fails to have a positive definite $\phi$-component of the rotation  Killing vector $\xi_\phi= \frac{\partial}{\partial \phi}$
for all values of the metric parameters, one must have $\ell \ge a $ which is different than the no-naked singularity condition  $|\ell m| \ge 2 a $. These two conditions not only restrict the maximum value of the spin, but they also restrict the minimum value of the mass to be $m \ge 2$. 

\begin{figure}[ht]

 \includegraphics[width=11cm]{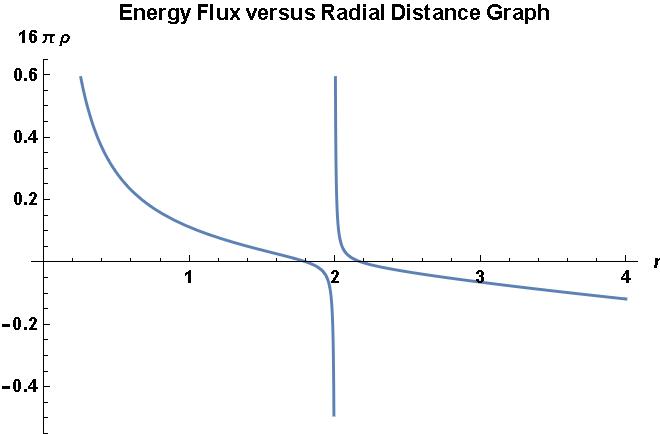}
\caption{ This figure represents the energy flux of the fluid dual to the rotating BTZ black hole. We chose the parameters as: $m=4, a=0.75, \ell=1$, hence the event horizon is at  $r_{\text{H}_{outer}}= 1.96317$, while $r_{\text{H}_{inner}}= 0.382035$. At the ergosphere radius, $r_{\text{ergosphere}}=2.00$, there is a discontinuity just like in the case of the four-dimensional Kerr black hole. The energy flux is continuous inside the ergosphere until it diverges at the center.}
\label{BTZEnergyfluxvsRadial}
\end{figure}

\begin{figure}[ht]
\includegraphics[width=11cm]{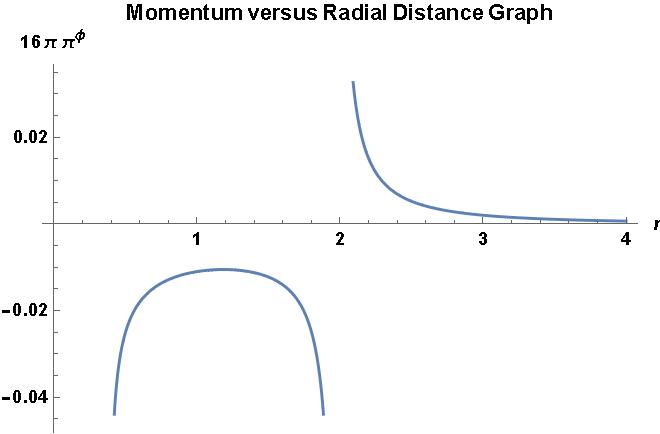}
\caption{{This figure represents the momentum density of the fluid dual to the rotating BTZ black hole. The parameters are $m=4, a=0.75, \ell=1$, the BTZ radius $r_{\text{H}_{outer}}= 1.96317$ while $r_{\text{H}_{inner}}= 0.382035$.} At the ergosphere radius, $r_{\text{ergosphere}}=2.00$, there is a discontinuity. The momentum is continuous and negative inside the ergosphere up until it diverges at the center, discontinuous at the ergosphere radius, and asymptotically zero at infinity, as expected. }
\label{BTZMomentumvsRadial}
\end{figure}

From now on, we shall assume $m > 2$ and $a< \ell$ so that one has $g_{\phi\phi}>0$ (so that close timelike curves are prohibited), such that the rotation Killing vector
is spacelike everywhere and the geometry has an event horizon. One further argument in using the Boyer-Lindquist form of the BTZ metric is the following: the recent work \cite{Arslaniev_original} provides the membrane construction of the Kerr black hole in the Boyer-Lindquist coordinates, and hence these coordinates are practical since we also analyze the BTZ black hole as a constant cross-section of the Kerr black hole. One naturally expects the membrane of the Kerr metric to be related to the membrane of the BTZ black hole in some sense. 
Moreover, in the previous section, we provided the static version of the black hole, and to find the transport coefficients, we will have to fix the bulk viscosity
$\zeta$ to its static counterpart. In these coordinates, one can write the metric in the form \cite{Kerrintroduction}:
\begin{equation}
    ds^2
     =-\left(F_tdt+\omega d\phi\right)^2+ F_r^2dr^2+\left(F_\phi^2+\omega^2\right)d\phi^2.
\end{equation}
Now, we should decompose this metric in the $(1+1+1)$-form:
\begin{equation}
    ds^2=\left(-u_\mu u_\nu+n_\mu n_\nu+\gamma_{\mu}\gamma_{\nu}\right)dx^\mu dx^\nu,
\end{equation}
where $u_\mu dx^\mu:=F_tdt+\omega d\phi$, $n_\mu dx^\mu:=F_r dr$, $\gamma_{\mu}dx^\mu:=\sqrt{F_\phi^2+\omega^2 }d\phi$.
As it should be clear, the BTZ metric is circularly symmetric, even though it is rotating, as the circular symmetry and rotation are compatible in 2+1 dimensions. This is somewhat consequential as all the metric functions are only $r$-dependent and there is no angular dependence. Therefore, one has a vanishing acceleration as in the static case:
\begin{equation}
    a_\nu=n^\gamma\nabla_\gamma n_\nu=0.
\end{equation}
The extrinsic curvature tensor of the timelike membrane, with a spacelike normal vector $n^\mu$, is given as $K_{\mu\nu}=h{^\gamma}_{\mu}\nabla_\gamma n_\nu$, and $K$ is its trace. They read explicitly as  
\begin{align}
    K_{\mu\nu}&=
\begin{pmatrix}
 -\frac{\sqrt{\Delta} \partial_r f}{2 r} & 0 & \frac{a \sqrt{\Delta} \partial_r f}{2 r} \\
 0 & 0 & 0 \\
 \frac{a \sqrt{\Delta} \partial_r f}{2 r} & 0 & \frac{\sqrt{\Delta} (2 r-a^2 \partial_r f)}{2 r} 
\end{pmatrix}, && K=\frac{ \left(r \partial_r f+2 f\right)}{2 \sqrt{\Delta}} .
\end{align}
The $1d$ cross-section of the extrinsic curvature can be calculated by taking the Lie derivative of $\gamma^\mu$ along the null vector $l^\mu$ 
\begin{equation}
    k_{\nu}=\left({\cal {L}}_l \gamma\right)_\nu=l^\mu\nabla_\mu\gamma_\nu+\gamma^\mu\nabla_\mu l_\nu 
    = \sqrt{\frac{\Delta}{f}}\left(\frac{ 2 r f^2-a^2 \partial_r f}{2 r^2 f }\right)\delta_\nu^\phi.
\end{equation}
The contraction of the  $1d$ cross-section and the extrinsic curvature gives the null expansion that we need:
\begin{equation}
    \Theta=\gamma^\mu k_\mu
    =\frac{2 r f^2 \Delta-a^2 \Delta \partial_r f}{2 a^2 r^2 f+2 r^4 f^2}.
    \label{nullexpansionrotating}
\end{equation}
By using the extrinsic curvature and the horizon metric, one can find the stress tensor of the stretched horizon as:
\begin{gather*}
    t^{\text{stretched}}_{\mu\nu}=\frac{1}{{2 r \sqrt{\Delta}}}
\begin{pmatrix}
  \left(a^2 \partial_r f-2 r f^2\right) & 0 & -a  \left(\left(a^2+r^2\right) \partial_r f-2 r \left(f-1\right) f\right) \\
 0 & 0 & 0 \\
 -a  \left(\left(a^2+r^2\right) \partial_r f-2 r \left(f-1\right) f\right) & 0 &  \left(a^2+r^2\right)^2 \partial_r f-2 a^2 r \left(f-1\right)^2 \\
\end{pmatrix}.
\end{gather*}
Note that the divergence-canceling term in the stress tensor at infinity, discussed above, does not contribute to the horizon.
 According to the membrane paradigm, the Newtonian viscous fluid description of the stress tensor should be identified with the stress tensor of the boundary-stretched horizon. The $tt$-components yield the energy density \cite{Arslaniev_original}:
\begin{equation}
\rho=\frac{1}{8 \pi }\left(\frac{\partial_r \log F_t^2}{2 F_r^2}-\frac{ \left(r \partial_r f+2 f\right)}{2 F_r \sqrt{\Delta}}\right)
=\frac{\left(r^2-\ell^2 m\right)^2-a^2 \ell^2}{8 \pi  \ell^2 r \left(\ell^2 m-r^2\right)},
    \label{energy}
\end{equation}
when $a\rightarrow 0$, we have the correct limit. The plot for the energy density can be found in Fig\eqref{BTZEnergyfluxvsRadial}
\begin{equation}
    \rho=\frac{1}{8 \pi r}\left(m-\frac{r^2}{\ell^2}\right).
\end{equation}
 As in the case of the Kerr black hole membrane paradigm, the static limit case should constrain the value of the bulk viscosity $\zeta$. In the static case, bulk viscosity vanishes as we have seen. Hence, we impose the same for the rotating case
\begin{equation}
    \zeta=0.
    \label{bulkviscosity}
\end{equation}
Furthermore, the fluid for the rotating black hole should have a momentum vector:
\begin{equation}
\pi^\phi=\frac{1}{16 \pi}\frac{\omega \partial_r \log \frac{\omega}{F_t}}{ F_r \left(F_\phi^2+\omega^2\right)}
    =\frac{a }{8\pi}\frac{1}{  \sqrt{r^2-\ell^2 m} \sqrt{a^2 \ell^2-\ell^2 m r^2+r^4}},
    \label{momentum}
\end{equation}
which vanishes when the static limit is achieved as expected. The plot for the momentum density is given in Fig\eqref{BTZMomentumvsRadial}

\begin{figure}[ht]%
\centering
\subfigure{%
\label{fig:first}%
\includegraphics[height=2in]{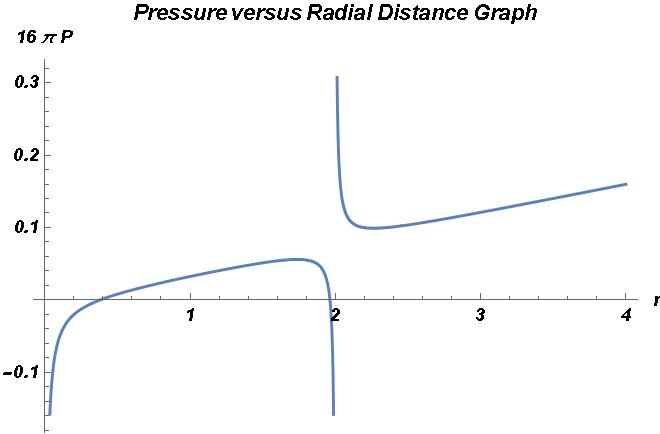}}%
\qquad
\subfigure{%
\label{fig:second}%
\includegraphics[height=2in]{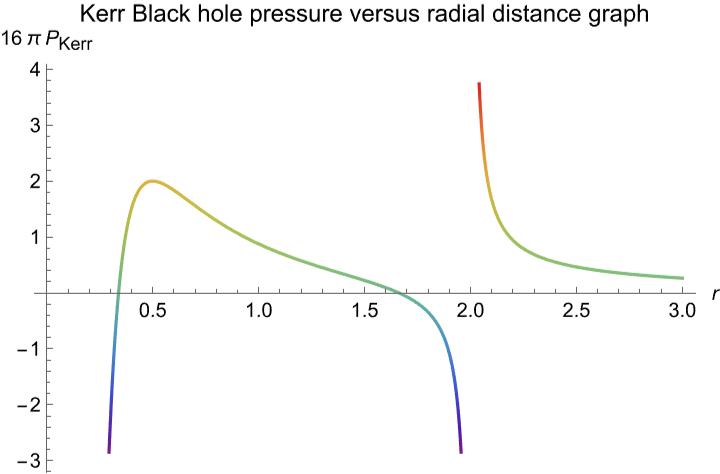}}%
\caption{{This figure represents the comparison between the BTZ and the Kerr black hole's dual fluid pressure. For the BTZ black hole we have chosen $m=4, a=0.75, \ell=1$, $r_{\text{H}_{outer}}= 1.96317$ while $r_{\text{H}_{inner}}= 0.382035$. For the Kerr black hole, we have chosen $m=1, a=0.75$, $r_{\text{H}_{outer}}= 1.66144$ while $r_{\text{H}_{inner}}= 0.338562$,  One can see that at the ergosphere radius,} $r_{\text{ergosphere}}=2.00$, there is a discontinuity in both pressure values and the behaviors of the pressure are similar.}
\label{BTZpressurevsradial}
\end{figure}

Also, there are no shearing effects for the BTZ black holes since it has a $1d$ space-like cross-section. Shear should be the traceless counterpart of the space-like cross-section. Since it is a vector, in $1d$, there is no such behavior of the fluid:
\begin{equation}
    \sigma^\mu=0.
\end{equation}
Now, we can look at the behavior of the pressure plotted as in the  Fig\eqref{BTZpressurevsradial}
\begin{equation}
    P=\frac{1}{8 \pi }\left(\frac{ \left(r \partial_r f+2 f\right)}{2F_r  \sqrt{\Delta}}-\frac{2 r f^2 \Delta-a^2 \Delta \partial_r f}{2 a^2 r^2 f+2 r^4 f^2}\right)=-\frac{a^2 \ell^2-\ell^2 m r^2+r^4}{8 \pi  \ell^4 m r-8 \pi  \ell^2 r^3},
    \label{pressure}
\end{equation}

To check whether this is correct, one should understand that in the limit $a\longrightarrow 0$, the surface gravity that is written as $P=\frac{\kappa}{8\pi}$ should correspond to that of the static BTZ black hole:
\begin{equation}
    \lim_{a\to0} P=\frac{1}{8 \pi}\frac{r}{ \ell^2} .
    \label{limitingpressure}
\end{equation}
On the horizon, the pressure is $P=\frac{1}{8 \pi}\frac{ \sqrt{m}}{\ell} $, which gives the correct value of surface gravity.

\section{Lorentz violating static and rotating spacetimes}\label{LorentzViolation}
Lorentz-violating gravity theories have been extensively studied with different approaches varying from string theory to Horava-Lifshitz gravity \cite{Kostelecky:1988zi,
Horava:2009uw,
Alfaro:2001rb,
Carroll:2001ws,
Yang:2023wtu}.  Even though Lorentz symmetry introduces an equivalence to all inertial reference frames in a given theory. One can introduce a spontaneously broken Lorentz symmetry into the theory through some non-minimal couplings to the graviton field. In \cite{Yang:2023wtu}, such a coupling exists through Kalb-Ramond 3-forms. This makes the theory explicitly Lorentz invariant; however, the vacuum-to-vacuum expectation value does not acquire such an invariance. In \cite{Yang:2023wtu}, there exists an exact solution which depends on this Lorentz-violating parameter $\ell$ (not to be confused with AdS radius in the previous section). We apply the Newman-Janis algorithm to produce its rotating counterpart and show the equivalence in the non-violating regime and non-rotating regime. The rotating solution to Bumblebee gravity with Lorentz violation is introduced in the paper. Still, to the best of our knowledge, the rotating solution to non-minimally coupled Einstein-Kalb-Ramond theory does not exist in the literature. After introducing the static and rotating black hole solutions for the theory mentioned above, we will apply the membrane paradigm to the black holes and find their transport coefficients.
Let us first give a brief review of how the Lorentz-violating parameter affects the form of the metric functions of the Schwarzschild and Kerr black holes. After that, we employ the paradigm machinery in the sense of \cite{Ulus} and \cite{Arslaniev_original}.
\subsection{Lorentz violating Schwarzschild spacetime}

Through the results of \cite{Yang:2023wtu}, we will introduce the Schwarzschild black hole, which depends on the Lorentz-violating parameter $\ell$
\begin{equation}
   ds^2= -f dt^2+f dr^2+r^2(d\theta^2+\sin^2\theta d\phi^2),
\end{equation}
where $f(r)=\frac{1}{1-\ell}-\frac{2 m}{r}$ where $m$ is the mass and $\ell$ is the Lorentz violating parameter. Due to observational tests, $\ell$ is constrained to be small. For this spacetime, the Kretschmann scalar is found to be:
\begin{equation}
    K=\frac{48 m^2}{r^6}-\frac{16 \ell m}{(1-\ell)r^5}+\frac{4 \ell^2}{(1-\ell)^2r^4},
\end{equation}
which reduces to that of the Schwarzschild spacetime at the 
 $\ell\rightarrow 0$. The event horizon is located at
\begin{equation}
    r_\text{H}=(1-\ell)2 m.
\end{equation}
\subsection{Lorentz violating Kerr spacetime}
Now, we will apply the Newman-Janis Algorithm. There is a non-trivial Lorentz violation, and we have to modify the effects of the seed function complexification. The metric is
\begin{equation}
    ds^2=-\left(\frac{1}{1-\ell}-\frac{2 m}{r}\right)dt^2+\left(\frac{1}{1-\ell}-\frac{2 m}{r}\right)^{-1}dr^2+r^2\left(d\theta^2+\sin^2\theta d\phi^2\right).
    \label{LorentzViolatingSchwarzschildMetric}
\end{equation}
The following Newman-Janis complexification algorithm can be applied
\begin{flalign}
    &r\to r-ia\sqrt{1-\ell}\cos\theta, \quad\quad a\to \sqrt{1-\ell} a,\quad\quad
    r\Bar{r}=\Sigma= r^2+(1-\ell)a^2\cos^2\theta.
\end{flalign}
Concerning these judiciously \footnote {For every hair we should add the violation parameter while considering the limits of hair charges, which go to zero, reducing to well-known forms of black holes.} chosen complexifications, we will be able to find the corresponding rotating black hole. One observes that the coefficient of $dtd\phi$ should be upgraded as $f(r,\theta)-\frac{1}{(1-\ell)}$ so that the cross-term can be reduced into a familiar Kerr counterpart. Through these definitions, one can find the $\Delta$ as:
\begin{flalign}
    \Delta=& \left(\frac{1}{1-\ell}-\frac{2 m r}{\Sigma}\right)\Sigma+a^2\sin^2\theta
    =\frac{r^2-2 m r+ (1-\ell)a^2}{1-\ell}.
    \label{deltaviolatingKerr}
\end{flalign}
Then the black hole becomes:
\begin{flalign}
    ds^2=&-\left(\frac{1}{1-\ell}-\frac{2 m r}{\Sigma}\right)dt^2+\frac{\Sigma}{\Delta}dr^2-2 a \sqrt{1-\ell}\frac{2 m r}{\Sigma}\sin^2\theta dt d\phi+\Sigma d\theta^2\notag\\&+\Big\{r^2+a^2(1-\ell)+\frac{2 m r}{\Sigma}(1-\ell)a^2\sin^2\theta\Big\}\sin^2\theta d\phi^2,
\end{flalign}
where $\Sigma=a^2 (1-\ell) \cos ^2(\theta )+r^2$, when $\ell\to 0$, it reduces to the Kerr black hole
\begin{equation*}
    ds^2=-(1-\frac{2 m r}{\Sigma})dt^2+\frac{\Sigma}{\Delta}dr^2-2 a \frac{2 m r}{\Sigma}\sin^2\theta dt d\phi+\Sigma d\theta^2+\Big\{r^2+a^2+\frac{2 m r}{\Sigma}a^2\sin^2\theta\Big\}\sin^2\theta d\phi^2.
\end{equation*}
 Also if $\ell\neq 0$ but $a=0$ we again find the Lorentz violating Schwarzschild black hole solution \eqref{LorentzViolatingSchwarzschildMetric}.
\subsection{The membrane paradigm and the Lorentz violation}
By choosing the metric seeds as:
\begin{align}
    &F_t=\sqrt{\frac{1}{1-\ell}-\frac{2 m r}{ \Sigma  }} ,
  \quad\quad F_r=\sqrt{\frac{ \Sigma  }{\Delta}},
    \quad\quad\omega=-\frac{a (1-\ell) (2 m r)}{ \Sigma  }F_t^{-1},\notag\\
    &F_\phi=\left[\frac{a^2 (1-\ell) \sin ^4\theta  (2 m r)}{ \Sigma  }+\sin^2\theta \left(a^2 (1-\ell)+r^2\right)\right]^{\frac{1}{2}}, 
\end{align}
with these definitions, the generic metric becomes:
\begin{equation}
    ds^2=-F_t^2dt^2-2\omega F_tdtd\phi+F^2_\phi d\phi^2+ F_r^2dr^2+\Sigma d\theta^2,
    \label{completedsquareformm}
\end{equation}
where \eqref{completedsquareformm} can be completed to a square \cite{Kerrintroduction}
\begin{align}
    ds^2&=-\left(F_tdt^2+\omega d\phi\right)^2+ F_r^2dr^2+\Sigma d\theta^2+\left(F_\phi^2+\omega\right)d\phi^2.
     \label{completedsquareform2}
\end{align}
Now, we should identify this metric with a $(2+1+1)$-dictionary:
\begin{equation}
    ds^2=\left(-u_\mu u_\nu+n_\mu n_\nu+\gamma_{AB}{e}{^A}_{\mu}{e}{^B}_{\nu}\right)dx^\mu dx^\nu.
\end{equation}
Let $u_\mu dx^\mu=F_tdt+\omega d\phi$, $n_\mu dx^\mu=F_r dr$, $\gamma_{\mu\nu}dx^\mu dx^\nu=\Sigma d\theta^2+\left(F_\phi^2+\omega^2\right)d\phi^2$
The structure of this metric can be put in the foliated form, such that one can directly start to declare the important factors that underlie the membrane paradigm of black holes:
\begin{flalign}
    K_{tt}&=\sqrt{\frac{\Delta}{\Sigma^5}}m\left(\Sigma-r\partial_r\Sigma\right),\quad\quad\quad\quad K_{t\phi}=-\sqrt{\frac{\Delta}{\Sigma^5}}a (\ell-1) m \left(\Sigma-r \partial_r \Sigma\right)\,\quad\quad\quad\quad
    K_{rr}=0,&&\notag\\
    K_{\theta\theta}&=\sqrt{\frac{\Delta}{\Sigma}}\frac{\partial_r \Sigma}{2},\quad\quad
    K_{\phi\phi}=\frac{1}{2}\sqrt{\frac{\Delta}{\Sigma^3}}\sin^2\theta\left(\Sigma(r^2-a^2(\ell-1))-2a^2(\ell-1)m r \sin^2\theta\right).&&
\end{flalign}
The trace of the extrinsic curvature reads:
\begin{flalign}
 K&= \frac{1}{A(r,\theta)}\left(-\frac{1}{2} a^2 (\ell-1)^2 m^2 r^2 (-4 \cos2\theta +\cos4\theta + 8 \ell -5) \partial_r \Sigma\notag\right.\\&\left. +\sin^2\theta \Sigma^2 \left(\left(a^2 (\ell-1)-r^2\right) \partial_r \Sigma -(\ell-1) m \left(a^2 \cos2\theta +a^2  (1- 2 \ell) +6 r^2\right)\right)\right.\\&\left.+a^2 (\ell-1)^2 m^2 r (-4 \cos2\theta +\cos4\theta + 8 \ell -5) \Sigma-2 r \sin^2\theta \Sigma^3\right)\notag,&&
\end{flalign}
where 
\begin{flalign*}
A(r,\theta)&=2 \Delta  \left(\frac{\Sigma}{\Delta }\right)^{3/2} \left(\sin^2\theta \Sigma \left(\left(a^2 (\ell-1)-r^2\right) \Sigma+(\ell-1) m r \left(-a^2 \cos2\theta\right.\right.\right.\\&\left. \left.\left.+a^2  (2\ell-1) -2 r^2\right)\right)+\frac{1}{2} a^2 (\ell-1)^2 m^2 r^2 (-4 \cos2\theta +\cos4\theta + 8 \ell -5)\right).&&
\end{flalign*}
where its functional form is given below:
\begin{equation}
   P=\frac{(\ell-1) m \Delta \left(\Sigma-r \partial_r \Sigma \right)}{8 \pi  \Sigma^2 (\Sigma+2 (\ell-1) m r)}.
\end{equation}
As one can see, the pressure of the Lorentz violating fluid perfectly matches with the usual Kerr fluid at the limit $\ell\mapsto 0$. One can see the rest of the transport coefficients from the Appendix (\ref{Lorentzviolatingappendix})
\begin{figure}[h]
\includegraphics[width=10cm]{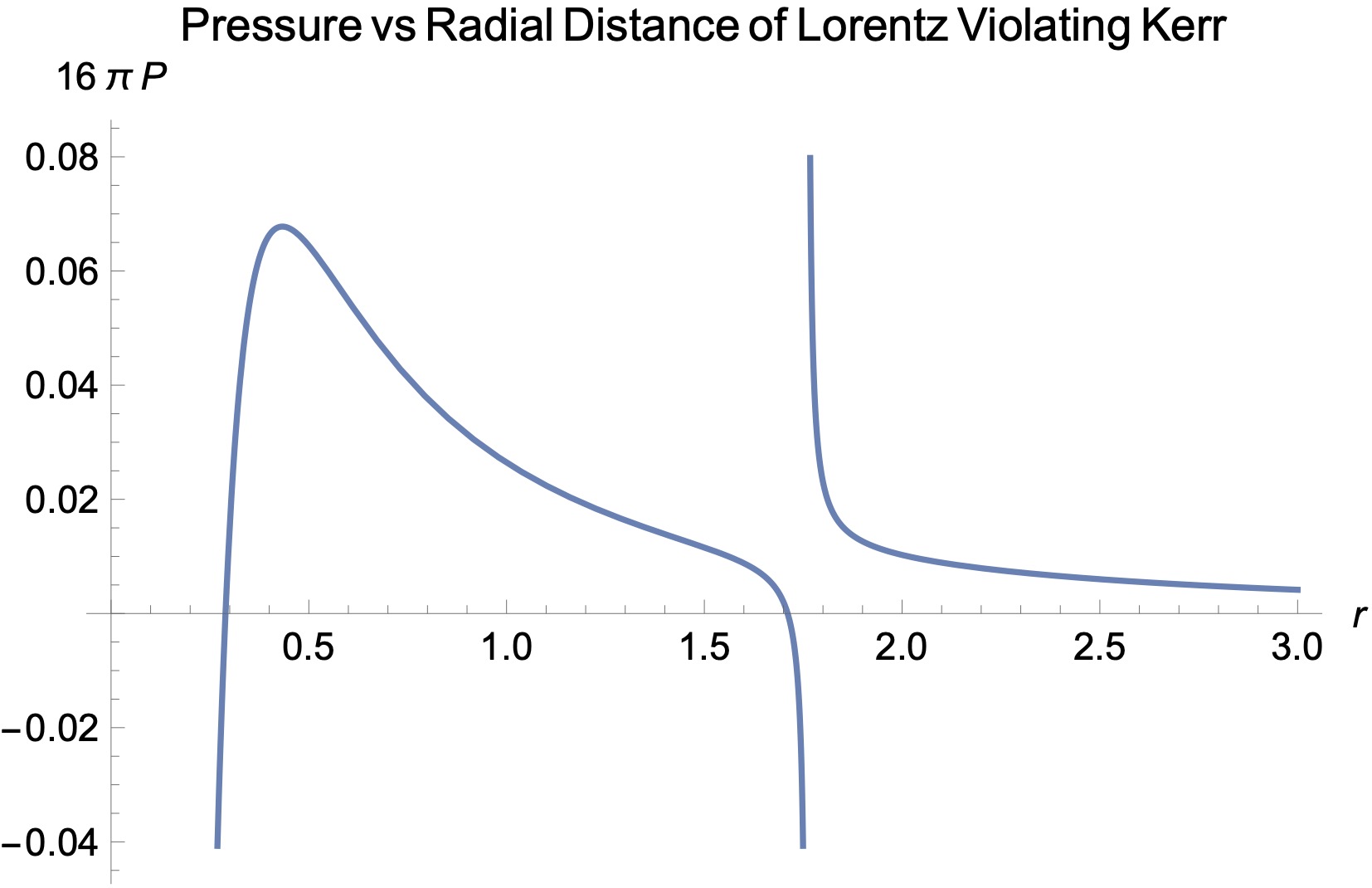}
\caption{This figure represents Lorentz violating Kerr black hole's dual fluid pressure values scaled with $16\pi$ when we choose $m=1, a=0.75, \ell=0.12$, horizon radius $r_{\text{H}_{outer}}= 1.71$ while $r_{\text{H}_{inner}}= 0.29$. One can see that at the ergosphere radius, $r_{\text{ergosphere}}=2$, there is a discontinuity just like in the case of the Kerr black hole. The pressure diverges negatively at the central singularity, discontinuous at the ergosphere radius at the equatorial plane, and is asymptotically zero at infinity while vanishing at the inner and outer event horizons.}
\label{pvsrLorentzviolating}
\end{figure}

\section{Asymptotically safe Schwarzschild spacetime}\label{asymptoticalsafetyschwarzschild}
    We now study the renormalization group (RG) inspired Schwarzschild-type metric \cite{Bonanno:2006eu}. Even though the membrane paradigm seems to be a classical theory of gravity, "asymptotic safety" ideas can be added as a classical correction to the Schwarzschild metric. From this point of view, quantum corrections are generically added by assuming a running Newton's coupling $G(k)$ in the specific form as a function of radial distance \cite{Bonanno:2000ep}.
    \begin{equation}
        G(r)=\frac{G_0 r^3}{r^3+\Tilde{\omega}G_0 [r+\gamma G_0 m]},
    \end{equation}
    where $\Tilde{\omega},\gamma$ are some constants.
which satisfies the asymptotic safety scenario \cite{Weinberg:1980gg}.
 We can distinguish the bare seed and the running seed from each other by the redefinition:
 \begin{equation}
     f(r)=1-\frac{2 G(r) m}{r}.
 \end{equation}
 The metric can be corrected with this newly defined seed such that we have the RG improved Schwarzschild black hole.
 RG improvement induces a quantum ergosphere such that it separates the event horizon and the apparent horizon from each other, which is not the case in classical Schwarzschild spacetime, in the Schwarzschild coordinates. This separation induces a critical mass value, resulting in an extremal limit that determines the final state of the evaporation, also known as the remnant. The new horizon radii, which accommodate the critical mass, are given as:
 \begin{equation}
     r_\pm= G_0 m\left(1\pm\sqrt{1-\Omega}\right),
 \end{equation}
 where $\Omega=\frac{m^2_{cr}}{m^2}$. By using similar ideas, one can find the transport coefficients of the RG-improved Schwarzschild black hole as:
\begin{align}
    \Theta=\frac{2}{r}f ,\quad\quad\sigma_{AB}=0 ,\quad\quad
    \kappa=\frac{\partial_r f}{2}.
\end{align}

\begin{align}
    \Theta=\frac{2}{r}\left(1-\frac{2 m}{r}\frac{G_0 r^3}{r^3+G_0 r}\right) ,& &\sigma_{AB}=0 ,&&
    \kappa=\frac{1}{2}\left(\frac{G}{r^2}-\frac{G'}{r}\right).
\end{align}
Now, if we try to find the value for the ongoing null fluxes:
\footnotesize
\begin{align}
\Theta = \frac{2}{G_0 m \left( 1 \pm \sqrt{1 - \Omega} \right)} 
\left( 
1 - \frac{2m}{G_0 m \left( 1 \pm \sqrt{1 - \Omega} \right)} 
\left( 
\frac{G_0 \left( G_0 m \left( 1 \pm \sqrt{1 - \Omega} \right) \right)^3}
{ \left( G_0 m \left( 1 \pm \sqrt{1 - \Omega} \right) \right)^3 + G_0 \left( G_0 m \left( 1 \pm \sqrt{1 - \Omega} \right) \right) } 
\right) 
\right).
\end{align}
\normalsize
As $m\rightarrow m_{cr}$, one has $\Theta<0$, which shows the decrease in the surface area. However, the behavior at $m=m_{cr}$ drastically changes. Unlike classical Schwarzschild black holes, there is a critical mass that gives a remnant. If we assume $\gamma= 1$ and $m_{cr}=m$ the radius of the remnant becomes $r_{rem}=G_0 m$. At this stage, the transport coefficients give the following results. 
\begin{flalign}
    \kappa(r_{rem})=0 \implies P=0,\quad\quad\quad\quad\quad
    \Theta(r_{rem})=\frac{2}{G_0 m}\left(\frac{1-G_0 m^2}{1+G_0 m^2}\right).
\end{flalign}
These results can be understood under the membrane paradigm. Zero pressure ensures that there are no non-zero external forces that give the final stage of the evaporation as a stable solution. However, the non-zero null expansion is generically unexpected for the Membrane paradigm approach. It shows that null geodesics are never tangential while we are going into the final stage of evaporation and predicts an absence of horizon for the remnant.
However, the Membrane paradigm, by its nature, works for near-horizon approaches. If we solve the null expansion such that there exists a null surface wrapping around the remnant:
\begin{flalign}
    \Theta(r_{rem})=0\notag\implies  m=\frac{1}{\sqrt{G_0}},
\end{flalign}
which is exactly 1 Planck mass. Hence, from the perspective of the membrane paradigm, the remnant should have a null surface wrapped around it. The theory predicts that even at the final stages of black hole evaporation, the event horizon is not lost. The applicability of the Newman-Janis algorithm to the RG improved black hole is non-unique since the complexification scheme works best for seed functions up to quadratic in $r$. Allowing the running Newton's constant induces higher orders of $r$, which makes the judicious choice of complexification non-trivial. Hence, we leave the rotating case for future work.

\section{A more general charged (non-)rotating black holes}\label{stringyblackholes}
\subsection{A static axion-dilaton black hole}
For this section, we study stringy generalizations beyond Einsteinian black holes, which are extensively studied in the literature. This constitutes solid checks for the viability of the paradigm, whether or not it works for higher global charges defined on the black hole. Moreover, we deduced that the paradigm brings seemingly different types of black holes into the same setting via its dual fluid representation. Working on the well-studied black holes reveals that this behavioral pattern is universal, which we will consider in a companion paper \cite{AgcaTekin:2025}.  
In \cite{Horne:1992zy, Gibbons:1987ps}, a generalization of a static black hole as a solution to some low-energy string theory. The line element reads
\begin{flalign}
    ds^2&=-f(r) dt^2+f(r)^{-1}dr^2+h(r)r^2(d\theta^2+\sin^2\theta d\phi^2).\notag\\
    f(r)&=1-\frac{ R_1}{r}(1-\frac{ R_2}{r})^{\frac{1-\alpha ^2}{\alpha ^2+1}}\quad\quad\quad h(r)=r \left(1-\frac{ R_2}{r}\right)^{\frac{\alpha ^2}{\alpha ^2+1}}
    \label{staticstringy}
\end{flalign}
where $ R_1$, $ R_2$ are horizon radii. For $\alpha= 0$ this solution reduces to the unique solution of Einstein-Maxwell theory, i.e, the Reisner-Nordstrom black hole.
Let us define a Parikh-Wilczek-like $(2+1+1)$ decomposition such that
\begin{align}
    U_\mu dx^\mu=f^{\frac{1}{2}}dt, && n_\mu dx^\mu=f^{-\frac{1}{2}}dr,&&    \gamma_{\mu\nu}=h\begin{pmatrix}
r^2 & 0 \\
0 & r^2\sin^2\theta 
\end{pmatrix},
\end{align}
where $U_\mu U^\mu=-1$ and $n^\mu n_\mu=1$, $\{A,B\}=\{\theta,\varphi\}$.
This means that a $3$-dimensional space-like surface with spherical topology is rescaled with the factor $h(r)=r \left(1-\frac{ R_2}{r}\right)^{\frac{\alpha ^2}{\alpha ^2+1}}$.
One can calculate the acceleration vector of the normal $n^\mu$ as:
\begin{equation}
    a_\nu=n^\mu\nabla_\mu n_\nu=0.
\end{equation}
There is no acceleration as in the Schwarzschild case.
The extrinsic curvature becomes:
\begin{flalign}
    K_{\mu\nu}&=\nabla_\mu n_\nu  =\frac{1}{2}\sqrt{f}\begin{pmatrix}
-\partial_r f & 0 & 0 & 0\\
0 & 0 & 0 & 0 \\
  0& 0  & r(r \partial_r h+2 h) &0\\
  0& 0  &0   &r\sin^2 \theta \left(r \partial_r h+2h\right)
\end{pmatrix}.
\end{flalign}
The scalar extrinsic curvature is the contraction:
\begin{equation}
K=\sqrt{f}\left(\partial_r \ln{h}+\frac{2}{r}\right)+\partial_r \ln{\sqrt{f}}
\end{equation}
while the energy-momentum tensor of the dual viscous fluid is:
\begin{flalign}
    t^{\text{viscous}}_{\mu\nu}&=f^{-\frac{1}{2}}\rho U_\mu U_\nu+\alpha^{-1}\gamma_{\mu A}\gamma_{\nu B}\big(P\gamma^{AB}-2\eta\sigma^{AB}-\zeta\Theta\gamma^{AB}\big)+\pi^A(\gamma_{\mu A}U_\nu+\gamma_{\nu B}U_\nu),&&
    \label{3.126}
\end{flalign}
where the transport coefficients of the fluid can be identified as:
\begin{align}
    &\rho=-\frac{1}{8\pi},  &\eta=\frac{1}{16\pi}&,
    &P=\frac{\kappa}{8\pi},  &&\zeta=-\frac{1}{16\pi},&
    &\pi^A=0.\notag
\end{align}
Recalling the stress tensor
    \begin{equation}
        t^{\text{stretched}}_{\mu\nu}=\frac{1}{16\pi f^{\frac{1}{2}}}\left(\gamma_{\mu\nu}\left(r h \partial_r f+f\left(2h+r\partial_r h\right)\right)-\left(\frac{2f}{r}+f\partial_r\ln{h}\right)U_\mu U_\nu\right),
    \end{equation}
   and comparing it with $t^{\text{stretched}}$.
    \begin{align}
        \Theta=\frac{2f}{r}+f\partial_r\ln{h},\quad\quad\sigma_{AB}=0,
        \quad\quad\kappa=\frac{1}{2}\partial_r f.
    \end{align}
Substituting the functions, one arrives at surface gravity, thus the pressure of the fluid,
 \begin{equation}
     \kappa=\frac{\left(1-\frac{ R_2}{r}\right){}^{\frac{2}{\alpha ^2+1}} \left(\left(\alpha ^2+1\right) r  R_1+ R_2 \left(\alpha ^2 (-r)+r-2  R_1\right)\right)}{2 \left(\alpha ^2+1\right) r \left(r- R_2\right){}^2},
 \end{equation}
 which $r\to  R_1$ reduces to
 \begin{equation}
     \kappa(r= R_1)=\frac{\left(1-\frac{ R_2}{ R_1}\right){}^{\frac{1-\alpha ^2}{\alpha ^2+1}}}{2  R_1}
     \label{alphadependentsurfacegravity}
 \end{equation}
in \eqref{alphadependentsurfacegravity} agreement with \cite{Gibbons:1987ps}. For the particular case of  $\alpha=\sqrt{3}$, one has the so-called GGHS-Schwarzschild black hole for which the surface gravity is.
\begin{equation}
    \kappa(r= R_1)=\frac{1}{2  R_1 \sqrt{1-\frac{ R_2}{ R_1}}}.
\end{equation}
If we set $R_1=2m-\frac{q^2}{m}$ and $R_2=\frac{q^2}{m}$, we have
\begin{equation}
    \kappa^{\text{GGHS-Sch}}=\frac{m}{2\sqrt{2}}\frac{1}{\sqrt{\left(2m^2-q^2\right)\left(m^2-q^2\right)}},
\end{equation}
which in the $q\rightarrow 0$ limit correctly reduces to the surface gravity of the Schwarzschild black hole.

\subsection{A rotating axion-dilaton black hole}
Some solutions of heterotic string theory correspond to black holes of Einstein-Maxwell-Dilaton-Axion gravity at the low-energy limits. The spherically symmetric solution, using the Newman-Janis method \cite{Yazadjiev:1999ce}, can be upgraded to a rotating one, of which the metric \cite{Horne:1992zy} is:
\begin{align}
    ds^2&=-fdt^2-2a \sin^2\theta \left(f-1\right)dt d\phi
    +\frac{\Sigma}{\Delta}dr^2+\Sigma d\theta^2\notag\\&+\sin^2\theta \left(-\left(a^2 \sin^2\theta \left(f-1\right)\right)+a^2+r \left(\frac{q^2}{m}+r\right)\right)d\phi^2.&&
\end{align}
 Choosing the metric seeds as:
\begin{flalign}
F_t&=\sqrt{f},\quad\quad\quad\quad\quad\quad
   F_r=\sqrt{\frac{\Sigma}{\Delta}},\quad\quad\quad\quad\omega=-a \sin^2\theta (1-f)F^{-1}_t,\notag\\
   F_\phi&=\sin\theta\left[ \left(a^2 \sin^2\theta \left(1-f\right)+\left(a^2+r \left(\frac{q^2}{m}+r\right)\right)\right)\right]^{\frac{1}{2}}, 
    \quad\quad f=1-\frac{2 m r}{ \Sigma},
\end{flalign}
where 
\begin{equation}
    \Delta=a^2+r \left(r-\left(2 m-\frac{q^2}{2 m}\right)\right),\quad\quad\quad
    \Sigma=a^2 \cos^2\theta+r \left(r-\left(2 m-\frac{q^2}{2 m}\right)\right)+2 m r,
\end{equation}
One can bring this metric into a form that is amenable to the membrane paradigm calculations. The components of the extrinsic curvature read as:
\begin{flalign}
    K_{tt}&=\frac{\Delta^{\frac{1}{2}}}{\Sigma^\frac{3}{2} }m (\Sigma-r \partial_r \Sigma),\notag
    &&K_{\phi\phi}=\frac{1}{2 }\sqrt{\frac{\Delta}{\Sigma}}\sin^2\theta \left(\frac{2 a^2 m r \sin^2\theta}{\Sigma}+a^2+r \left(\frac{q^2}{m}+r\right)\right),\notag
    &&K_{rr}=0,\notag\\
   K_{\theta\theta}&=\frac{1}{2 }\sqrt{\frac{\Delta}{\Sigma}}\partial_r \Sigma,&
  &K_{t\phi}=\frac{\Delta^{\frac{1}{2}}}{\Sigma^\frac{3}{2} }a m \sin^2\theta \left(\Sigma-r \partial_r \Sigma\right),
\end{flalign}
and its trace becomes:
\begin{equation*}
  K=\frac{1}{A(r,\theta)}\left(m (a^2+r^2)+q^2 r\right) \partial_r \Sigma+(2 m r+q^2) \Sigma-m\left(a^2 m \cos 2\theta\\+a^2 m+6 m r^2+4 q^2 r\right),
\end{equation*}
where 
\begin{equation*}
    A(r,\theta)=2 \sqrt{\frac{\Sigma}{\Delta}} \left(\Sigma \left(a^2 m+r \left(m r+q^2\right)\right)-m r \left(a^2 m \cos 2\theta+a^2 m+2 r \left(m r+q^2\right)\right)\right).
\end{equation*}
The pressure reads as follows:
\begin{flalign}
    P&= \frac{1}{Y(r,\theta)}\left[ m \left(a^2+r \left(\frac{q^2}{2 m}-2 m+r\right)\right) \left(a^2 \cos ^2\theta +\frac{q^2 r}{2 m}-r \left(\frac{q^2}{2 m}+2 r\right)+r^2 \right) \right],
\end{flalign}
where
\begin{align*}
    Y(r,\theta)&= \left(-a^2 \cos ^2\theta -\frac{q^2 r}{2 m}+2 m r-r^2\right) \left(a^2 \cos ^2\theta +\frac{q^2 r}{2 m}+r^2\right)^2. &&
\end{align*}
Limiting to the non-rotating regime, one can fix the bulk viscosity as $\zeta=-\frac{1}{16\pi}$ \cite{Arslaniev_original}. One can see the rest of the transport coefficients in Appendix (\ref{stringysolutionappendix}).
 
\subsection{Membrane description of the $4$-dimensional black hole from the Kaluza-Klein reduction in $5$-dimensions}
One can dimensionally reduce a particular solution of $5$-dimensional vacuum Einstein gravity to a 4-dimensional boosted, translationally invariant black hole with a non-trivial dilaton field \cite{Horne:1992zy}. Here, we find the transport coefficients of this black hole, and we show that in the non-rotating limit $a\rightarrow 0$, it reproduces the boosted Schwarzschild black hole with a dilaton field such that the bulk viscosity of the rotating black hole can be fixed through the non-rotating counterpart.

\begin{align}
    ds^2=&-\frac{1-Z}{B}dt^2-\frac{2a Z \sin^2\theta}{B\sqrt{1-v^2}} dt d\phi
    +\frac{B \Sigma}{\Delta}dr^2+B \Sigma d\theta^2\\&+\sin^2\theta \left(B(r^2+a^2)+a^2\sin^2\theta \frac{Z}{B}\right)\sin^2\theta d\phi^2,&&\notag
\end{align}
where
\begin{equation}
    B=\sqrt{1+\frac{v^2 Z}{1-v^2}},\quad\quad Z=\frac{2 m r}{\Sigma},
    \quad\quad\Delta= r^2+a^2-2 m r , \quad\quad\Sigma=r^2+a^2\cos^2\theta,
     \label{metricfunctions}
\end{equation}
where $ m$ is the mass and $a$ is the angular momentum parameter, $v$ is the degree of boost parameter. The functions in the Parikh-Wilczek-like splitting are read:
\begin{flalign*}
f&=\frac{1-Z}{B},
    \quad\quad\quad\quad\quad\quad F_t=\sqrt{f},\quad\quad\quad\quad\quad\quad
   F_r=\sqrt{\frac{B \Sigma}{\Delta}},
   \end{flalign*}
\begin{flalign}
    \quad\quad\quad F_\phi&= \left(\sin^2\theta\left(a^2+r^2\right) B+\frac{a^2 \sin^4\theta Z}{B}\right)^{\frac{1}{2}},
   & \omega=-\frac{1}{\sqrt{1-v^2}BF_t}\left(a\sin^2\theta-B F^2_t\right),&&
\end{flalign}
The components of the extrinsic curvature read as
\begin{flalign}
    K_{tt}&=\frac{1}{2}B^\frac{3}{2} \sqrt{\frac{ \Delta}{\Sigma}}\left( B \partial_r Z-(Z-1) \partial_r B\right),\notag&
    K_{t\phi}=\Delta^\frac{1}{2}\frac{a \sin^2\theta \left(B \partial_r Z-Z \partial_r B\right)}{2 \sqrt{1-v^2}  \sqrt{B^3 \Sigma{}}},\notag&&
     K_{rr}=0,\notag&&
\end{flalign}
\begin{flalign}
     K_{\theta\theta}&=\frac{1}{2}\sqrt{\frac{\Delta}{B \Sigma}}\partial_r \Sigma,&
    K_{\phi\phi}&=\frac{1}{2}\sqrt{\frac{\Delta}{B \Sigma}}\left(\left(a^2+r^2\right) \sin^2\theta B+\frac{a^2 \sin ^4 \theta Z}{B}\right),&&&&&
\end{flalign}
while its trace reads:
\begin{flalign}
  K=&\frac{1}{A(r,\theta)}\left(v^2-1\right) B^3 \left(\Sigma \left(\left(a^2+r^2\right) \partial_r Z+2 r \left(Z-1\right)\right)\notag\right.\\&\left.+\left(a^2+r^2\right) \left(Z-1\right) \partial_r \Sigma\right)+a^2 \sin^2\theta B \left(\Sigma \left(2 v^2 Z-v^2+1\right) \partial_r Z\right.\\&\left.+Z \left(v^2 Z-v^2+1\right) \partial_r \Sigma\right)+2 a^2 \sin^2\theta \Sigma Z \left(v^2 (-Z)+v^2-1\right) \partial_r B,\notag
\end{flalign}
where
\begin{equation}
A(r,\theta)=2 \Delta \left(\frac{B \Sigma}{\Delta}\right)^{3/2} \left(\left(v^2-1\right) \left(a^2+r^2\right) B^2 (Z-1)+a^2 \sin^2\theta Z \left(v^2 Z-v^2+1\right)\right).
\end{equation}
The pressure reads as follows:
\begin{flalign}
   P=&\frac{\Delta \left(B \partial_r Z-(Z-1) \partial_r B\right)}{16 \pi  B^2 \Sigma (Z-1)}.
   \label{pressurekaluza}
\end{flalign}
Using the non-rotating limit, one can fix the bulk viscosity as $\zeta=-\frac{1}{16\pi}$. Then the pressure reads:
\begin{flalign}
    P&=\frac{1}{Y(r,\theta)}m \left(a^2+r \left(r-2 m\right)\right) \left(r^2-a^2 \cos^2 \theta\right) \Big(a^2 \left(v^2-2\right) \cos^2 \theta\\&+r \left(v^2 (r-2 m)-2 r\right)\Big),\notag
\end{flalign}
where
\begin{flalign}
    Y(r,\theta)&=16 \pi  \left(a^2 \cos^2 \theta+r^2\right)^2 \left(a^2 \cos^2 \theta+r (r-2 m)\right) \left(a^2 \left(v^2-1\right) \cos^2 \theta\right.\\&\left.+r \left(v^2 (r-2 m)-r\right)\right) \left(1-\frac{2 m r v^2}{\left(v^2-1\right) \left(a^2 \cos^2 \theta+r^2\right)}\right)^\frac{1}{2}.\notag
\end{flalign}
The behavior of the function can be analyzed on the equatorial plane $\theta=\frac{\pi}{2}$:
\begin{equation*}
P =\left\{
        \begin{array}{ll}
            \infty &  r=0,\\
            
            0&  r=\infty,\\
             0 & r=r_{\text{Kerr}}=m+m\left(1-\frac{a^2}{m^2}\right)^\frac{1}{2},\\
            \infty & r=r_{\text{ergo}}=2m.
        \end{array}
    \right..
\end{equation*}
Moreover, in the limit $\theta\to\frac{\pi}{2},a\to 0$, we should get the boosted Schwarzschild solution. 
\begin{figure}[h]
\includegraphics[width=10cm]{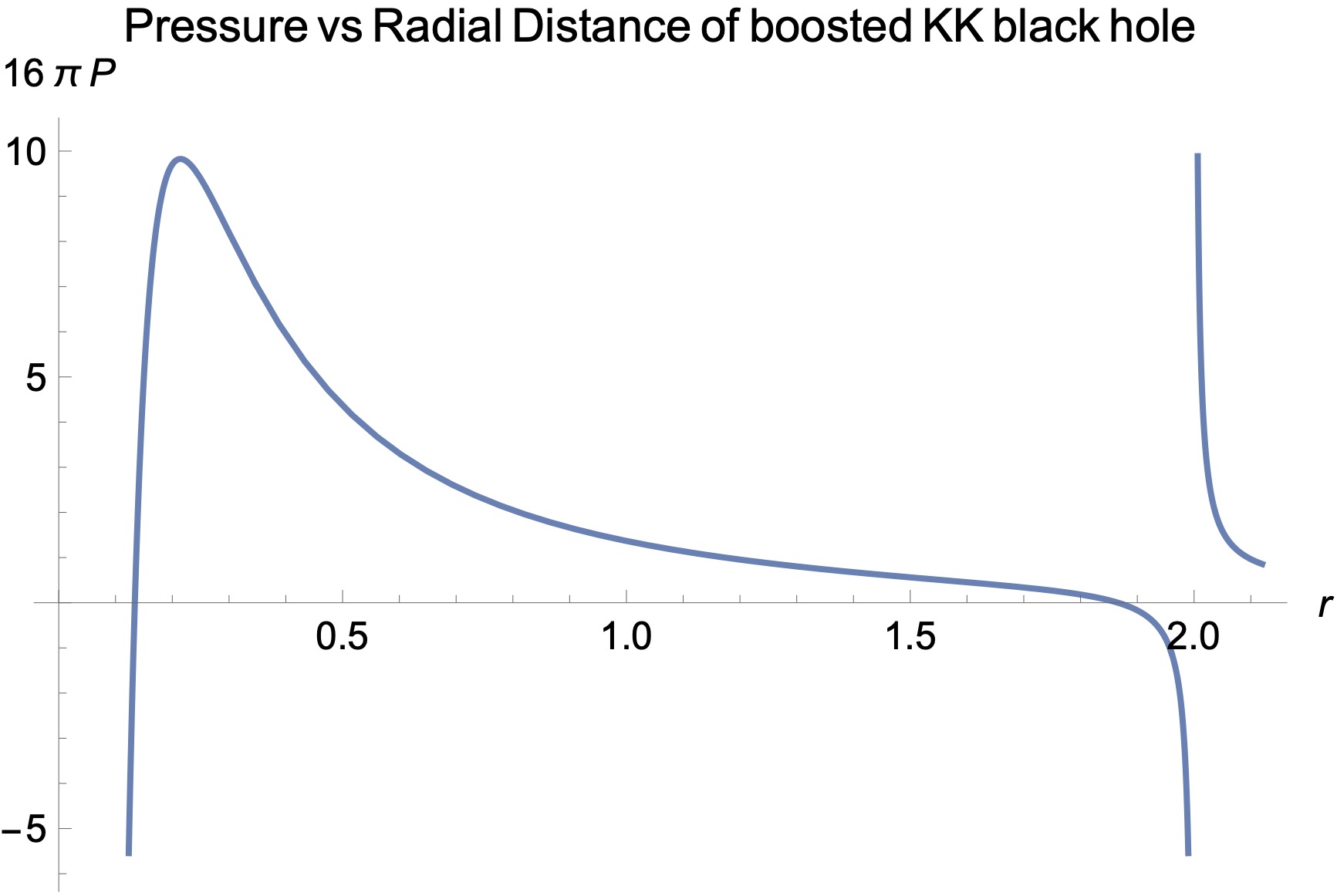}
\caption{This figure represents boosted KK black hole's dual fluid pressure values scaled with $16\pi$ when we choose $m=1, a=0.75, v=0.25$, horizon radius $r_{\text{H}_{outer}}= 1.87$ while $r_{\text{H}_{inner}}= 0.13$. One can see that at the ergosphere radius, $r_{\text{ergosphere}}=2 $ there is a discontinuity just like in the case of the Kerr black hole. The pressure diverges negatively at the central singularity, discontinuous at the ergosphere radius at the equatorial plane, and is asymptotically zero at infinity while vanishing at the inner and outer event horizons.}
\label{KKboostedplot}
\end{figure}
The horizon radius is not affected by the boost, hence $ r_\text{H}=2m$ gives:
\begin{equation}
    P(a=0,r=2m)=\frac{1}{8 \pi }\frac{\sqrt{1-v^2}}{ 4 m},
    \label{boostedsurfacegravityy}
\end{equation}
where  $\kappa=\frac{\sqrt{1-v^2}}{ 4 m}$.
 One can find the rest of the transport coefficients in Appendix (\ref{Kaluzakleinsolutionappendix}).

\section{Conclusions }
We studied several black holes within the effective framework of the so-called fluid-membrane. Among the black holes studied here, the $2+1-$dimensional BTZ black hole is somewhat different, with a two-dimensional null horizon and a one-dimensional spatial cross-section. It is well-known that $2+1$ dimensional General Relativity, with or without a cosmological constant, is trivial with no local bulk degrees of freedom. However, the theory admits static and rotating black holes if the cosmological constant is negative, despite its local triviality, that is, locally the BTZ black hole is $AdS_3$, yet globally it has an event horizon and a non-trivial causal structure like the four-dimensional black holes. There has been a plethora of papers on the BTZ black hole since its first description \cite{Banados:1992wn} for apparent reasons, as one can study some of the quantum behavior of black holes without the complications of bulk gravitons. The BTZ black hole has been studied within the context of black hole chemistry \cite{Kubiznak:2016qmn, Hajian:2021hje}  where a volume and a pressure are also assigned to the solution in addition to the mass $m$ and angular momentum parameter $a$. Pressure of the black hole in these works arises as a dual thermodynamical coordinate to the cosmological constant. However, the membrane paradigm naturally suggests a definition of pressure that does not solely rely on the cosmological constant, and when the pressure is analytically continued to the rest of the black hole spacetime, it hints at the existence of a generalized van der Waals-type fluid description of the black hole spacetime \cite{AgcaTekin:2025}. Here, we gave a dual formulation of both static and rotating BTZ black holes in terms of a fluid membrane. We constructed the effective theory using the action formulation of Parikh and Wilczek. Much is known about the asymptotic symmetries and the expected $2d$ conformal field theory of $AdS_3$. It would be interesting to understand better the connections between the conformal field theory description $2d$ and the fluid-membrane description presented here \cite{Brown:1986nw}. We leave that for future work.
 
Using the Newman-Janis algorithm, we found the rotating counterpart of
a previously known black hole with a Lorentz-violating parameter. This parameter can be interpreted as an additional charge or hair of the black hole. The membrane description of this black hole gave interesting results:
There seems to be a shift given by the degree of the Lorentz violation in the radius of the region where astrophysical jets are possibly generated. Therefore, the degree of Lorentz violation $\ell$ can affect the process of jet generation in the case where the Lorentz-violating parameter is generated by a Kalb-Ramond field.
The membrane paradigm can also be used for quantum-corrected black holes, and in this vein, we have applied it to the case of asymptotic safety-inspired black holes, which yielded some non-trivial results. For example, in the renormalization group improved Schwarzschild black hole, in which Newton's constant depends on the radial distance, the membrane description of the evaporating black hole will predict the existence of a remnant with a 1 Planck mass endowed with an event horizon wrapped around it.

Furthermore, the membrane paradigm successfully generates the well-known properties of stringy (non-)rotating black holes with dilatonic and axionic charges, which constitutes a non-trivial test of the paradigm for non-Einsteinian black holes. Let us reiterate that the membrane paradigm is an effective theory that maps all properties of black holes to a constant negative energy density, constant shear, bulk viscosity, and incompressible fluid, which satisfies some solution of the Damour-Navier-Stokes equations. From this perspective, the membrane paradigm, as in the case of holographic fluid, should be able to capture the quasi-normal modes of a given black hole or compact objects \cite{Saketh:2024ojw} in terms of hydrodynamic modes. Two types of modes for hydrodynamics come from the explicit expansions of the component energy-momentum tensor \cite{Springer:2008js,Policastro:2002se}, or from the linearization of the Navier-Stokes equations through plane-wave solutions, one can arrive at both of the frequencies such as the shear mode, $\omega_{\text{shear}}(k)=-i\frac{\eta}{\rho+P}k^2$ which is a purely imaginary damping mode or the sound mode, $\omega_{\text{sound}}(k)=u_s k-\frac{i}{2}\frac{1}{\rho+P}(\frac{d-1}{d}+\frac{\zeta}{2\eta})k^2$, where $d$ is the dimension of the spacetime and $u_s$ is the speed of sound. Solving for the sound mode relies on relaxing the incompressibility condition and allowing for longitudinal waves. Most of the time, the fluid membrane is taken as an incompressible fluid; in any case, we note two situations to have a deeper understanding of the fictitious fluid with negative bulk viscosity. Let us first discuss the shear mode of membrane fluid. The quasi-normal modes corresponding to the shear mode of the fluid should have a negative value of the $\text{Im}(\omega_{\text{shear}}(k))<0$ for stability. The membrane paradigm unavoidably fixes the values for shear, bulk viscosities; however, the pressure depends on the radial distance as a function of surface gravity. The reader should remember that we analytically continue the pressure inside the black hole to be able to check how the external observer assigns the properties of the fluid instead of the real black hole. Moreover, zeros of the pressure will give us important black hole regions such as the ergoregion, the inner and outer event horizons, and the true singularity. By inserting the numerical values, one can find stable and unstable fluid regions naively by taking these as the quasi-normal modes through this prescription. It seems the membrane paradigm suggests that when $|\rho|>P(r)$, we have an unstable fluid; however when $|\rho|<P(r)$, we have a stable fluid. If we were to take the membrane paradigm fluid as a compressible fluid, it is a naive but interesting assumption that the transport coefficients keep their values exactly. The imaginary part of the sound mode should again have the values $\text{Im}(\omega_{\text{sound}}(k))<0$ to have a stable compressible fluid or quasi-normal modes for the black holes. In the expression $\text{Im}(\omega_{\text{sound}}(k))=-\frac{1}{2(\rho+P)}\left(\frac{(d-1)2\eta+d\zeta}{2d\eta}\right)$, the term inside the parenthesis is strictly bigger than zero for all dimensions.\footnote{One should not forget that for smaller than three dimensions bulk viscosity vanishes identically and otherwise it is negative.} This analysis shows that all rotating black holes under the paradigm approach behave like the subcritical temperature behavior of a specific van der Waals fluid, which corresponds to a vapor-liquid state. These issues will follow in \cite{AgcaTekin:2025}.

\section{Acknowledgments}
The authors thank Prof. Dr. Sakir Erkoc for their support of this work.
\appendix
\section{A brief review of Newman Janis algorithm}\label{appendix}
Let us write the steps of the Newman-Janis Algorithm. We have no claim of novelty on the basic construction of the algorithm.
\begin{enumerate}
    \item Spherically symmetric metrics have a \emph{seed function}. One can make the seed function visible by transforming to the Eddington-Finkelstein coordinates $(t,r,\theta,\phi)\longrightarrow (u,r,\theta,\phi)$
    \item Complexify both the seed function and the corresponding coordinates $f(r)\longrightarrow f(r,\Bar{r})$, find the Newman-Penrose null tetrads.
    \item Transform the metric to the Boyer-Lindquist coordinates.
\end{enumerate}
Let us start with a generic, static, spherically symmetric spacetime with the Schwarzschild-type seed metric \cite{Chou:2020nja}.
\begin{equation}
    ds^2=-f(r)dt^2+f^{-1}(r)dr^2+r^2d\Omega^2_2
\end{equation}
The Gaussian curvature $\kappa$ of $d\Omega^2_2=d\theta^2+K(\theta)d\phi^2$ can be fixed by choice:
\begin{equation*}
    \kappa=\begin{Bmatrix}
+1\; \text{if}\; S^2 \\
-1 \; \text{if} \; H^2
\end{Bmatrix}\implies K(\theta)=\begin{Bmatrix}
\sin\theta \; \text{if} \; \kappa=1 \\
\sinh\theta\; \text{if}\; \kappa=-1
\end{Bmatrix}.
\end{equation*}
Let $\kappa=1$ such that we have a spherical topology on $\mathcal{M}=\Sigma\otimes S^2$. Then the metric is
\begin{equation}
    ds^2=-f(r)dt^2+f^{-1}(r)dr^2+r^2(d\theta^2+\sin^2\theta d\phi^2),
\end{equation}
with the coordinates $x^\mu=\{t,r,\theta,\phi\}$. Transforming this into Eddington-Finkelstein coordinates gives:
\begin{align*}
    du=dt+f^{-1}(r)dr \implies & dt=du-f^{-1}(r)dr,  &
\end{align*}
such that $dt^2=du^2-2f^{-1}(r)dudr+(f^{-1}(r))^2dr^2$ transforms the metric to
\begin{flalign}
    ds^2&=-f(r)du^2-2dudr+r^2d\Omega^2_2 .
\end{flalign}
Now, let us pick a complex null tetrad. This null tetrad should obey the Newman-Penrose (NP) formalism \cite{Newman:1961qr}. If one can fix $\phi=\phi_0$ and $\theta=\theta_0$ such that $ds^2=0$:\footnote{We are considering the vanishing $(t,r)$ cross-section of the metric at constant spherical angles to find the null tetrads in $(t,r)$ directions}
\begin{flalign}
    -f(r)du^2-2dudr=0\quad\quad\quad\quad
    du= -\frac{f(r)}{2}dr .
    \label{2.24}
\end{flalign}
Hence, if one picks the set of null basis as $(l^a,n^a,m^a,\Bar{m}^a)$ where $\Bar{m}^a=(m^a)^*$. This choice should obey the following rules:
\begin{flalign}
  m^a\bar{m}^a=+1 ,\quad\quad\quad l^a n_a=-1 , \notag\\
   l^a l_a=n^a n_a=m^a m_a=\bar{m}^a \bar{m}_a=0 ,\notag\\
    l^a m_a=l^a \bar{m}_a=n^a m_a=n^a \bar{m}_a=0.
\end{flalign}
In the NP formalism, the tetrad basis consists of two real basis vectors $\{l^a,n^a\}$ and two complex basis vectors $\{m^a,\Bar{m}^a\}$. They are self-orthogonal and cross-orthogonal, while real and complex subsets are normalized to $\{-1,1\}$ respectively in the $(-,+,+,+)$ signature. A real subset of the basis can be seen from \eqref{2.24} easily.
\begin{align}
    l^a={\delta}{_r}^{a} , & & n^a={\delta}{_u}^{a}-\frac{f(r)}{2}{\delta}{_r}^{a}  ,
\end{align}
the imaginary part can be detected by choosing $t=t_0$ and $r=r_0$ where $ds^2=0$.
such that:
\begin{align}
    m^a=\frac{1}{\sqrt{2}r}({\delta}{_\theta}^{a}+\frac{i}{\sin\theta}{\delta}{_\phi}^{a} ), & &\Bar{m}^a=\frac{1}{\sqrt{2}r}({\delta}{_\theta}^{a}-\frac{i}{\sin\theta}{\delta}{_\phi}^{a} ).
\end{align}
In general, one can also find the null-tetrad by finding 
 the orthonormal tetrad in $(t,r,\theta,\phi)$ coordinates. For spherically symmetric metrics:
\begin{equation}
    ds^2=g^{\mu\nu}\partial_\mu\partial_\nu=\eta^{ab}{e}{_a}^{\mu}{e}{_b}^{\nu}\partial_\mu\partial_\nu
    =\frac{-1}{f(r)}\partial_t^2+f(r)\partial_r^2+\frac{1}{r^2}\partial_\theta^2+\frac{1}{r^2\sin^2\theta}\partial_\phi^2.
\end{equation}
One can choose the vielbeins in such a way that the metric can be orthogonalized
\begin{equation}
    {e}{_a}^{t}=-\sqrt{\frac{1}{f(r)} ,}\quad\quad {e}{_a}^{r}=\sqrt{f(r)},\quad\quad 
    {e}{_a}^{\theta}=\frac{1}{r} ,\quad\quad {e}{_a}^{\phi}=\frac{1}{r\sin\theta} .
    \label{2.27}
\end{equation}
such that the basis can be a linear combination of  \eqref{2.27}
\begin{equation}
   l_a = \frac{ {e}{_a}^{t}+ {e}{_a}^{r}}{\sqrt{2}},\quad\quad n_a=\frac{ {e}{_a}^{t}- {e}{_a}^{r}}{\sqrt{2}},\quad\quad
    m_a=\frac{ {e}{_a}^{\theta}+i {e}{_a}^{\phi}}{\sqrt{2}},\quad\quad \Bar{m}_a=\frac{ {e}{_a}^{\theta}-i {e}{_a}^{\phi}}{\sqrt{2}}.
    \label{2.28}
\end{equation}
The main part of the formalism is to complexify the $(r,u)$ subspace \cite{Erbin:2016lzq}. Let $r\longrightarrow \Tilde{r}-ia\cos\theta$, $u\longrightarrow\Tilde{u}+ia\cos\theta$,$\theta=\Tilde{\theta}$, and $\phi=\Tilde{\phi}$. The seed function $f(r)\in\mathbb{R}$ is naturally generalized into $f(r,\Bar{r})\in\mathbb{C}$, also the null-tetrad adapts into new coordinates. The transformation can be written as $\Tilde{Z}^a =\frac{\partial \Tilde{x}^\mu}{\partial x^\nu}Z^a$ where $Z^a=\{l^a,n^a,m^a,\Bar{m}^a\}$. The real part of the tetrad analytically continued to complex space.
\begin{align}
    \Tilde{l}^a= {\delta}{^a}_{r}, & & \Tilde{n}^a= {\delta}{^a}_{\Tilde{u}}-\frac{f(r,\Bar{r})}{2} {\delta}{^a}_{r}.
\end{align}
The complex part is trickier, but it boils down to a straightforward algebraic calculation.
\begin{equation}
    m^a=\frac{1}{\sqrt{2}(\Tilde{r}+ia\cos\theta)}\left( {\delta}{^a}_{\theta}+\frac{i}{\sin\theta} {\delta}{^a}_{\phi}-ia\left( {\delta}{^a}_{\Tilde{u}}- {\delta}{^a}_{\Tilde{r}}\right)\sin\theta\right) 
\end{equation}
For convenience, we will drop the "tilde" from now on. Moreover, the seed function should also be complexified.
\begin{equation}
    r\longrightarrow r-ia\cos\theta ,\quad\quad\quad\Bar{r}\longrightarrow r+ia\cos\theta ,\quad\quad\quad
    |r|^2=r\bar r=r^2+a^2\cos^2\theta:=\Sigma(r,\theta) .
     \label{complexification}
\end{equation}
The seed function reads as:
\begin{equation}
    f(r,\bar r)=1-\frac{2 M(r)}{r}\left(\frac{r^2}{\Sigma(r,\theta)}\right) ,
    \label{seedfunction}
\end{equation}
such that, $M(r)$ might include mass $m$, charge $q$ or cosmological constant $\Lambda$.
\begin{equation}
    M(r)=m-\frac{q^2}{2r}+\frac{\Lambda r^3}{6} .
    \label{2.32}
\end{equation}
The line element can be written as
\begin{flalign}
ds^2&=f(r,\theta)du^2+2dudr+2a\sin^2\theta\left[1-f(r,\theta)\right]dud\phi-2a\sin^2\theta drd\phi\notag\\&-\Sigma(r,\theta)d\theta^2-\sin^2\theta\left[(r^2+a^2)+a^2\sin^2\theta\left[1-f(r,\theta)\right]\right]d\phi^2.
\end{flalign}
This line element is not finalized; we can gauge away $g_{r\phi}$ and $g_{u\phi}$ components by an appropriate coordinate transformation. Let us transform our metric into Boyer-Lindquist type such that it becomes \emph{Hamilton-Jacobi Separable} \cite{chaotic2}, whereas, in the spherically symmetric form, it is also \emph{Klein-Gordon separable} \cite{Konoplya:2021slg}. One has to find a transformation $(u,r,\theta,\phi)\longrightarrow (t,r,\theta,\phi)$:
\begin{equation}
    du=dt+g(r)dr ,\quad\quad\quad
    d\phi=d\phi+h(r)dr.
\end{equation}
Hence, the line element becomes:
\begin{flalign}
    ds^2&=f(r,\theta)\left(dt+g(r)dr\right)^2+2(dt+g(r)dr)dr\notag\\&+2a\sin^2\theta\left(1-f(r,\theta)\right)(dt+g(r)dr)(d\phi+h(r)dr)\notag\\&-2a\sin^2\theta dr(d\phi+h(r)dr)-\Sigma(r,\theta)d\theta^2\notag\\&-\sin^2\theta\left((r^2+a^2)+a^2\sin^2\theta(1-f(r,\theta))\right)(d\phi+h(r)dr)^2 .
    \label{2.34}
\end{flalign}
If $g_{rt}=g_{r\phi}=0$, one can uniquely solve $g(r)$ and $h(r)$ as:
\begin{flalign}
    g(r)=&\frac{r^2+a^2}{f(r,\theta)\Sigma(r,\theta)+a^2\sin^2\theta} ,\quad\quad\quad
    h(r)=\frac{-a}{f(r,\theta)\Sigma(r,\theta)+a^2\sin^2\theta} .\label{2.36}
\end{flalign}
For simplicity, let the denominator be $\Delta(r,\theta)\equiv f(r,\theta)\Sigma(r,\theta)+a^2\sin^2\theta$, which is called the \emph{discriminant} for Kerr-like metrics in the Boyer-Lindquist coordinates. By substituting \eqref{2.36} into \eqref{2.34}, the metric can be simplified.
\begin{flalign}
    ds^2=&-f(r,\theta)dt^2+\frac{\Sigma(r,\theta)}{\Delta(r,\theta)}dr^2+2a(f(r,\theta)-1)\sin^2\theta dtd\phi+\Sigma(r,\theta)d\theta^2\\&+\left(r^2+a^2+(1-f(r,\theta))a^2\sin^2\theta\right)\sin^2\theta d\phi^2\notag
    \label{NJAgeneralrotatingblackhole}
\end{flalign}
\section{Some important transport coefficients for various theories}\label{transportcoefificients}
\subsection{Lorentz violating spacetime}\label{Lorentzviolatingappendix}
The components of the shear tensor become
\small
\begin{align}
\sigma_{\theta\theta} &= \frac{1}{B(r,\theta)}\left(\Delta  \left(-16 a^2 (\ell-1)^3 m^3 r^3 \left(\sin ^4\theta +\ell-1 \right) \partial_r \Sigma +4 (\ell-1)^2 m^2 r^2 \Sigma \left(\left(-4 a^2 \sin ^4\theta\notag\right.\right.\right.\right.\\&\notag\left.\left.\left.\left.+\sin ^2\theta  \left(-a^2\ell+a^2+r^2\right)-3 a^2 (\ell-1)\right) \partial_r \Sigma +2 a^2 (\ell-1) m\left(\sin ^4\theta +\ell-1 \right)\right)\right.\right.\\&\left.\left.+\sin ^2\theta  \Sigma^3 \left(\left(r^2-a^2 (\ell-1)\right) \partial_r \Sigma +2 (\ell-1) m \left(a^2 \sin ^2\theta -4 r^2\right)\right)+4 (\ell-1) m r \Sigma^2 \left(  \left(-a^2\sin ^4\theta \notag\right.\right.\right.\right.\\&\left.\left.\left.\left.-a^2\sin ^2\theta \ell+a^2\sin ^2\theta+r^2\sin ^2\theta\right) \partial_r \Sigma +2 (\ell-1) m \left(a^2 \sin ^4\theta +a^2 (\ell-1)-r^2 \sin ^2\theta \right)\right)\right.\right.\\&\left.\left.-2 r \sin ^2\theta  \Sigma^4\right)\right) \notag&&
\end{align}

\begin{flalign}
\sigma_{\phi\phi} &= \frac{1}{C(r,\theta)}\left(\Delta  \left(-8 a^2 (\ell-1)^4 m^3 r^2 \left(r \partial_r \Sigma -\Sigma\right)+8 a^2 (\ell-1)^3 m^2 r^2 \partial_r \Sigma  (\Sigma+2 (\ell-1) m r)\notag\right.\right.\\&\left.\left.+\partial_r \Sigma  (-\Sigma-2 (\ell-1) m r) \left(-\sin ^2\theta  (-\Sigma-2 (\ell-1) m r) \left(\left(r^2-a^2 (\ell-1)\right) \Sigma\right.\right.\right.\right.\\&\left.\left.\left.\left.-2 a^2 (\ell-1) m r \sin ^2\theta \right)-4 a^2 (\ell-1)^3 m^2 r^2\right)+2 \sin ^2\theta  (\Sigma+2 (\ell-1) m r)^2 \left(a^2 (\ell-1) m r \sin ^2\theta  \partial_r \Sigma\right.\right.\right.\notag\\&\left. \left.\left.-a^2 (\ell-1) m \sin ^2\theta  \Sigma+r \Sigma^2\right)-8 a^2 (\ell-1)^3 m^2 r \Sigma (\Sigma+2 (\ell-1) m r)\right)\right). \notag&&
\end{flalign}
\normalsize

where
\begin{flalign}
B(r,\theta)&=4 \Sigma^2 (\Sigma+2 (\ell-1) m r)^2 \left(\frac{2 a^2 (\ell-1) m r \left(\sin^4\theta  (-\Sigma)-2 (\ell-1) m r \left(\sin ^4\theta +\ell-1 \right)\right)}{\Sigma (\Sigma+2 (\ell-1) m r)}\notag\right.\\&\left.+\sin ^2\theta  \left(r^2-a^2 (\ell-1)\right)\right),\quad\quad\quad
C(r,\theta)=4 \Sigma^3 (\Sigma+2 (\ell-1) m r)^2.
\end{flalign}
And the non-zero momentum will become:
\begin{flalign}
  \pi^\phi&=\frac{1}{G(r,\theta)}\Bigg(a (\ell-1) m \Big(\Delta  \left(r \partial_r \Sigma  (3 \Sigma+4 (\ell-1) m r)-2 \Sigma (\Sigma+(\ell-1) m r)\right)\notag\\&-r \Sigma \partial_r \Delta (\Sigma+2 (\ell-1) m r)\Big)\Bigg),
\end{flalign}
where
\begin{flalign}
    G(r,\theta)&= 16 \pi  \Delta ^2 \left(\frac{\Sigma}{\Delta }\right)^{3/2} \sqrt{\frac{1}{(1-\ell)}-\frac{2 m r}{\Sigma}} \left(\sin ^2\theta  \Sigma \left(\left(a^2 (\ell-1)-r^2\right) \Sigma+(\ell-1) m r \left(-a^2 \cos2\theta\right.\right.\right.\notag\\&\left.\left.\left.+a^2  (2\ell-1) -2 r^2\right)\right)+\frac{1}{2} a^2 (\ell-1)^2 m^2 r^2 (-4 \cos2\theta+\cos4\theta+ 8 \ell -5)\right).
\end{flalign}
The null expansion of the black hole is:  
 \begin{flalign}
   \Theta&=\frac{\Delta}{N(r,\theta)}  \left(2 \left(-4 a^2 (\ell-1)^4 m^3 r^2 \left(r \partial_r \Sigma -\Sigma\right)+4 a^2 (\ell-1)^3 m^2 r^2 \partial_r \Sigma  (\Sigma+2 (\ell-1) m r)\right.\right.\notag\\&\left.\left.+\sin ^2\theta  (\Sigma+2 (\ell-1) m r)^2 \left(a^2 (\ell-1) m r \sin ^2\theta  \partial_r \Sigma -a^2 (\ell-1) m \sin ^2\theta  \Sigma+r \Sigma^2\right)\right.\right.\\&\left.\left.\notag-4 a^2 (\ell-1)^3 m^2 r \Sigma (\Sigma+2 (\ell-1) m r)\right)+\frac{\partial_r\Sigma (\Sigma+2 (\ell-1) m r)^2}{(\Sigma+2 (\ell-1) m r)} \left(2 a^2 (\ell-1) m r \left(\sin ^4\theta  (-\Sigma)\right.\right.\right.\notag\\&\left.\left.\left.-2 (\ell-1) m r \left(\sin ^4\theta +\ell-1 \right)\right)+(\Sigma+2 (\ell-1) m r)\sin ^2\theta  \left(r^2-a^2 (\ell-1)\right) \Sigma\right) \right)\notag,&&
\end{flalign}
where
\begin{flalign}
    N(r,\theta)&=\frac{2 \Sigma^2(\Sigma+2 (\ell-1) m r)^2}{\Sigma+2 (\ell-1) m r} \left(2 a^2 (\ell-1) m r \left(\sin ^4\theta  (-\Sigma)-2 (\ell-1) m r \left(\sin ^4\theta +\ell-1 \right)\right)\right.\notag\\&\left.+\left(\Sigma+2 (\ell-1) m r\right)\sin ^2\theta  \left(r^2-a^2 (\ell-1)\right) \Sigma\right).&&
\end{flalign}
\subsection{Stringy Solution}\label{stringysolutionappendix}
 The shear tensor becomes:
 \begin{flalign}
  \sigma_{\theta\theta}&=\frac{1}{D(r,\theta)}\left(\Delta \left(2 a^2 m^3 r^2 \partial_r \Sigma+2 a^2 m^3 r^2 \cos 2\theta \partial_r \Sigma-2 a^2 m^2 r \Sigma \partial_r \Sigma\right.\right.\notag\\&\left.\left.-2 a^2 m^2 r \cos 2\theta \Sigma \partial_r \Sigma+a^2 m \Sigma^2 \partial_r \Sigma-a^2 m^2 \Sigma^2+a^2 m^2 \cos 2\theta \Sigma^2\right.\right.\notag\\&\left.\left.+4 m^3 r^4 \partial_r \Sigma+4 m^2 q^2 r^3 \partial_r \Sigma-4 m^2 r^3 \Sigma \partial_r \Sigma-4 m q^2 r^2 \Sigma \partial_r \Sigma\right.\right.\\&\left.\left.+m r^2 \Sigma^2 \partial_r \Sigma+q^2 r \Sigma^2 \partial_r \Sigma-8 m^3 r^3 \Sigma-4 m^2 q^2 r^2 \Sigma+8 m^2 r^2 \Sigma^2\right.\right.\notag\\&\left.\left.+4 m q^2 r \Sigma^2-2 m r \Sigma^3-q^2 \Sigma^3\right)\right),\notag&&.
\end{flalign}
\begin{flalign}
\sigma_{\phi\phi}&=\frac{1}{C(r,\theta)}\left(\sin^2\theta \Delta \left(\partial_r \Sigma \left(-\Sigma^2 \left(a^2 m+r \left(m r+q^2\right)\right)+2 m r \Sigma \left(a^2 m \cos 2\theta\right.\right.\right.\right.\notag\\&\left.\left.\left.\left.+a^2 m+2 r \left(m r+q^2\right)\right)-2 m^2 r^2 \left(a^2 m \cos 2\theta+a^2 m+2 r \left(m r+Q^2\right)\right)\right)\right.\right.\\&\left.\left.+\Sigma \left(-m \Sigma \left(a^2 m \cos 2\theta-a^2 m+4 r (2 m r+q^2)\right)+(2 m r+q^2) \Sigma^2\right.\right.\right.\notag\\&\left.\left.\left.+4 m^2 r^2 (2 m r+q^2)\right)\right)\right),\notag&&
\end{flalign}
where
\begin{flalign}
  D(r,\theta)&=4 \Sigma (2 m r-\Sigma) (-a^2 m \Sigma-m r^2 \Sigma-q^2 r \Sigma+a^2 m^2 r \cos 2\theta\\&+a^2 m^2 r+2 m^2 r^3+2 m q^2 r^2),\quad\quad\quad
  C(r,\theta)=4 m \Sigma^2 (\Sigma-2 m r)^2.&&
\end{flalign}
The non-zero component of the  momentum will become:
\small
\begin{equation}
  \pi^\phi= \frac{1}{G(r,\theta)}\left(a m^2 \sqrt{\frac{\Sigma}{\Delta}} \left(r \Sigma \left(2 m r-\Sigma\right) \partial_r \Delta+\Delta \left(r \left(3 \Sigma-4 m r\right) \partial_r \Sigma+2 \Sigma \left(m r-\Sigma\right)\right)\right)\right).
\end{equation}
\normalsize
The null-expansion of the charged axion-dilaton black hole is:
\begin{flalign}
    \Theta&=\frac{1}{N(r,\theta)}\left(\Delta \left(\partial_r \Sigma \left(\Sigma \left(a^2 m+r \left(m r+q^2\right)\right) \left(\Sigma-4 m r\right)+2 m^2 r^2 \left(a^2 m \cos 2\theta\right.\right.\right.\right.\notag\\&\left.\left.\left.\left.+a^2 m+2 r \left(m r+q^2\right)\right)\right)+\Sigma \left(2 m \Sigma \left(a^2 m \sin^2\theta-2 r \left(2 m r+q^2\right)\right)+\left(2 m r+q^2\right) \Sigma^2\right.\right.\right.\\&\left.\left.\left.+4 m^2 r^2 \left(2 m r+q^2\right)\right)\right)\right)\notag,
\end{flalign}
where
\begin{flalign}
    G(r,\theta)&=\Sigma^3 \sqrt{1-\frac{2 m r}{\Sigma}} \left(\Sigma \left(a^2 m+r \left(m r+q^2\right)\right)-m r \left(a^2 m \cos 2\theta+a^2 q+2 r \left(m r+q^2\right)\right)\right)\notag,&&\\
N(r,\theta)&=2 \Sigma^2 \left(2 m r-\Sigma\right) \left(\left(m r \left(a^2 m \cos 2\theta+a^2 m+2 r \left(m r+q^2\right)\right)\notag\right.\right.\\&\left.\left.-\Sigma \left(a^2 m+r \left(m r+q^2\right)\right)\right)\right).
\end{flalign}
\subsection{Kaluza-Klein solution}\label{Kaluzakleinsolutionappendix}
 The shear tensor becomes:
 \begin{flalign}
  \sigma_{\theta\theta}&=\frac{\Delta}{4 B}\left(\frac{\partial_r \Sigma}{\Sigma}\right)-\frac{1}{C(r,\theta)}B \sin^2\theta \left(\left(a^2+r^2\right) \partial_r B+\frac{a^2 \sin^2\theta Z \left(v^2 (-Z)+v^2-1\right) \partial_r B}{\left(v^2-1\right) B^2 (Z-1)}\notag\right.\\&\left.+\frac{a^2 \sin^2\theta (v Z-v-1) (v Z-v+1) \partial_r Z}{\left(v^2-1\right) B (Z-1)^2}+2 r B\right).
\end{flalign}
\begin{flalign}
\sigma_{\phi\phi}&=\frac{\sin^2\theta \Delta}{4 B \Sigma^2} \left(\left(a^2+r^2\right) \Sigma \partial_r B+B \left(2 r \Sigma-\left(a^2+r^2\right) \partial_r \Sigma\right)\notag\right.\\&\left.+\frac{1}{L(r,\theta)}a^2 \sin^2\theta \Sigma Z \left(v^2 (-Z)+v^2-1\right) \partial_r B\right.\\&\left.+\frac{a^2 \sin^2\theta}{\left(v^2-1\right) B (Z-1)^2} \left(\Sigma (v Z-v-1) (v Z-v+1) \partial_r Z-(Z-1) Z \left(v^2 Z-v^2+1\right) \partial_r \Sigma\right)\right),\notag
\end{flalign}
where
\small
\begin{equation}
     C(r,\theta)=\left(a^2+r^2\right) \sin^2\theta B^2+a^2 \sin ^4 \theta Z \left(\frac{Z}{\left(v^2-1\right) (Z-1)}+1\right),\quad\quad
     L(r,\theta)=\left(v^2-1\right) B^2 (Z-1).
\end{equation}
\normalsize
The non-zero component of the  momentum will become:
\begin{flalign}
  \pi^\phi&=\frac{1}{G[r,\theta]}a \sqrt{1-v^2} \left(\Sigma \left(\Delta B (Z-2) \partial_r Z+(Z-1) Z \left(B \partial_r \Delta-2 \Delta \partial_r B\right)\right)\notag\right.\\&\left.-\Delta B (Z-1) Z \partial_r \Sigma\right),
\end{flalign}
where
\begin{flalign}
    G(r,\theta)&=  32 \pi  \Delta^2 \sqrt{\frac{1-Z}{B}} \left(\frac{B \Sigma}{\Delta}\right)^{3/2} \Big(\left(v^2-1\right) \left(a^2+r^2\right) B^2 (Z-1)\notag\\&+a^2 \sin^2\theta Z \left(v^2 Z-v^2+1\right)\Big).
\end{flalign}
The null-expansion of the charged axion-dilaton black hole is:
\begin{flalign}
    \Theta&=\frac{\Delta}{2 B \Sigma^2} \frac{\sin^2\theta \Sigma}{D(r,\theta)} \left(\left(\left(a^2+r^2\right) \partial_r B+\frac{a^2 \sin^2\theta Z \left(v^2 (-Z)+v^2-1\right) \partial_r B}{\left(v^2-1\right) B^2 \left(Z-1\right)}\notag\right.\right.\\&\left. \left.+\frac{a^2 \sin^2\theta (v Z-v-1) (v Z-v+1) \partial_r Z}{\left(v^2-1\right) B (Z-1)^2}+2 r B\right)+\partial_r \Sigma \right),
\end{flalign}
where
\begin{equation}
    D(r,\theta)=\left(a^2+r^2\right) \sin^2\theta B+\frac{a^2 \sin ^4 \theta Z}{B} \left(\frac{Z}{\left(v^2-1\right) (Z-1)}+1\right).
\end{equation}


\begin{thebibliography}{99}

\bibitem{Ulus}
C.~U.~Agca and B.~Tekin,
 Membrane paradigm approach to the Johannsen-Psaltis black hole,
Phys. Rev. D \textbf{109}, no.10, 10 (2024).

\bibitem{Johannsen:2011dh}
T.~Johannsen and D.~Psaltis,
A Metric for Rapidly Spinning Black Holes Suitable for Strong-Field Tests of the No-Hair Theorem,
Phys. Rev. D \textbf{83}, 124015 (2011).

\bibitem{Finkelstienoriginal}
D.~Finkelstein,
Past-future asymmetry of the gravitational field of a point particle,
Phys. Rev. \textbf{110}, 965-967 (1958).

\bibitem{membrane_THORNE}
K.~S.~Thorne, R.~H.~Price and D.~A.~Macdonald,
\textit{Black Holes: The Membrane Paradigm},
(Yale University, Press, London, 1986)

\bibitem{Damour_Actio_membranes}
T.~Damour,
Quelques proprietes mecaniques, electromagnetiques, thermodynamiques et quantiques des trous noirs (these de doctorat d etat)
, 1979 (unpublished)




\bibitem{MembraneHorizonsBlackHolesNewClothes}
M.~K.~Parikh,
Membrane horizons: The black hole's new clothes,
arXiv:hep-th/9907002.



\bibitem{Parikh_original}
M.~Parikh and F.~Wilczek,
An action for black hole membranes,
Phys. Rev. D \textbf{58}, 064011 (1998).

\bibitem{Silvestrini:2025lbe}
M.~Silvestrini, E.~Maggio, S.~Chakraborty and P.~Pani,
One Membrane to Love them all: Tidal deformations of compact objects from the membrane paradigm,
[arXiv:2506.16516 [gr-qc]].


\bibitem{Bah:2020ogh}
I.~Bah and P.~Heidmann,
Topological Stars and Black Holes,
Phys. Rev. Lett. \textbf{126}, no.15, 151101 (2021).


\bibitem{Banados:1992wn}
M.~Banados, C.~Teitelboim and J.~Zanelli,
The black hole in three-dimensional space-time,
Phys. Rev. Lett. \textbf{69}, 1849-1851 (1992).




\bibitem{Carlip:1995qv}
S.~Carlip,
The (2+1)-dimensional black hole,
Class. Quant. Grav. \textbf{12}, 2853-2880 (1995).



\bibitem{Arslaniev_original}
A.~J.~Nurmagambetov and A.~M.~Arslanaliev,
Kerr black holes within the membrane paradigm,
LHEP \textbf{2022}, 328 (2022).

\bibitem{Arslanaliev:2023vev}
A.~M.~Arslanaliev and A.~J.~Nurmagambetov,
Taking the Null-Hypersurface Limit in the Parikh-Wilczek Membrane Approach,
East Eur. J. Phys. \textbf{2024}, no.4, 35-50 (2024).

\bibitem{agca_2023}
C.~U.~Agca,
An analysis on the membrane paradigm of black holes,
, M.S.\textendash{}Master of Science. Middle East Technical University, 2023 (unpublished)

\bibitem{Israel_junction_condition}
D.~Marolf and S.~Yaida,
Energy conditions and junction conditions,
Phys. Rev. D \textbf{72}, 044016 (2005).


\bibitem{MembraneView}
R.~H.~Price and K.~S.~Thorne,
Membrane viewpoint on black holes: Properties and evolution of the stretched horizon,
Phys. Rev. D \textbf{33}, 915 (1986).


\bibitem{Balasubramanian:1999re}
V.~Balasubramanian and P.~Kraus,
A stress tensor for Anti-de Sitter gravity,
Commun. Math. Phys. \textbf{208}, 413 (1999).


\bibitem{stokeshypothesis}
M.~Papalexandris,
On the applicability of Stokes' hypothesis to low Mach-number flows,
Continuum Mechanics and Thermodynamics \textbf{32}, (2019).


\bibitem{Prasia:2016esx}
P.~Prasia and V.~C.~Kuriakose,
 Quasinormal modes and thermodynamics of linearly charged BTZ black holes in massive gravity in (anti) de Sitter space-time,
Eur. Phys. J. C \textbf{77}, 27 (2017)



\bibitem{surfacegravityBTZ}
B.~Gwak and B.~H.~Lee,
Thermodynamics of three-dimensional black holes via charged particle absorption,
Phys. Lett. B \textbf{755}, 324 (2016).




\bibitem{Erbin:2016lzq}
H.~Erbin,
Janis-Newman algorithm: generating rotating and NUT charged black holes,
Universe \textbf{3}, no.1, 19 (2017).



\bibitem{Kim:1998iw}
H.~Kim,
Spinning BTZ black hole versus Kerr black hole: A closer look,
Phys. Rev. D \textbf{59}, 064002 (1999).


\bibitem{Kerrintroduction}
M.~Visser,
The Kerr Spacetime: A Brief Introduction,
Kerr Fest: Black Holes in astrophysics, general relativity and quantum gravity,
 arXiv:0706.0622.

\bibitem{Kostelecky:1988zi}
V.~A.~Kostelecky and S.~Samuel,
Spontaneous Breaking of Lorentz Symmetry in String Theory,
Phys. Rev. D \textbf{39}, 683 (1989)

\bibitem{Horava:2009uw}
P.~Horava,
Quantum Gravity at a Lifshitz Point,
Phys. Rev. D \textbf{79}, 084008 (2009).

\bibitem{Alfaro:2001rb}
J.~Alfaro, H.~A.~Morales-Tecotl and L.~F.~Urrutia,
Loop quantum gravity and light propagation,
Phys. Rev. D \textbf{65}, 103509 (2002).
\bibitem{Carroll:2001ws}
S.~M.~Carroll, J.~A.~Harvey, V.~A.~Kostelecky, C.~D.~Lane and T.~Okamoto,
Noncommutative field theory and Lorentz violation,
Phys. Rev. Lett. \textbf{87}, 141601 (2001).

\bibitem{Yang:2023wtu}
K.~Yang, Y.~Z.~Chen, Z.~Q.~Duan and J.~Y.~Zhao,
Static and spherically symmetric black holes in gravity with a background Kalb-Ramond field,
Phys. Rev. D \textbf{108}, no.12, 124004 (2023).



\bibitem{Bonanno:2006eu}
A.~Bonanno and M.~Reuter,
Spacetime structure of an evaporating black hole in quantum gravity,
Phys. Rev. D \textbf{73}, 083005 (2006).

\bibitem{Bonanno:2000ep}
A.~Bonanno and M.~Reuter,
Renormalization group improved black hole space-times,
Phys. Rev. D \textbf{62}, 043008 (2000).

\bibitem{Weinberg:1980gg}
S.~Weinberg,
Ultraviolet divergences in quantum theories of gravitation, General Relativity: An Einstein centenary survey. Cambridge University Press. pp. 790-831.


\bibitem{AgcaTekin:2025}
C.~U.~Agca and B.~Tekin,
Black Holes as Multicomponent van der Waals Fluids: A Universal Behavior,
to appear (2025).



\bibitem{Horne:1992zy}
J.~H.~Horne and G.~T.~Horowitz,
Rotating dilaton black holes,
Phys. Rev. D \textbf{46}, 1340-1346 (1992).

\bibitem{Gibbons:1987ps}
G.~W.~Gibbons and K.~I.~Maeda,
Black Holes and Membranes in Higher Dimensional Theories with Dilaton Fields,
Nucl. Phys. B \textbf{298}, 741-775 (1988).


\bibitem{Yazadjiev:1999ce}
S.~Yazadjiev,
Newman-Janis method and rotating dilaton axion black hole,
Gen. Rel. Grav. \textbf{32}, 2345-2352 (2000).


\bibitem{Kubiznak:2016qmn}
D.~Kubiznak, R.~B.~Mann and M.~Teo,
Black hole chemistry: thermodynamics with Lambda,
Class. Quant. Grav. \textbf{34}, no.6, 063001 (2017).

\bibitem{Hajian:2021hje}
K.~Hajian, H.~{\"O}z{\c{s}}ahin and B.~Tekin,
First law of black hole thermodynamics and Smarr formula with a cosmological constant,
Phys. Rev. D \textbf{104}, no.4, 044024 (2021).




\bibitem{Brown:1986nw}
J.~D.~Brown and M.~Henneaux,
Central Charges in the Canonical Realization of Asymptotic Symmetries: An Example from Three-Dimensional Gravity,
Commun. Math. Phys. \textbf{104}, 207-226 (1986)


\bibitem{Saketh:2024ojw}
M.~V.~S.~Saketh and E.~Maggio,
Quasinormal modes of slowly-spinning horizonless compact objects,
Phys. Rev. D \textbf{110}, no.8, 084038 (2024).


\bibitem{Springer:2008js}
T.~Springer,
Sound Mode Hydrodynamics from Bulk Scalar Fields,
Phys. Rev. D \textbf{79}, 046003 (2009).


\bibitem{Policastro:2002se}
G.~Policastro, D.~T.~Son and A.~O.~Starinets,
From AdS / CFT correspondence to hydrodynamics,
JHEP \textbf{09}, 043 (2002).


\bibitem{Chou:2020nja}
Y.~C.~Chou,
Extension Rules of Newman--Janis Algorithm for Rotation Metrics in General Relativity,
Phys. Sci. Int. J. \textbf{24}, no.6, 1--14 (2020).


\bibitem{Newman:1961qr}
E.~Newman and R.~Penrose,
An approach to gravitational radiation by a method of spin coefficients,
J. Math. Phys. \textbf{3}, 566-578 (1962).


\bibitem{chaotic2}
O.~Zelenka and G.~Lukes-Gerakopoulos,
Chaotic motion in the Johannsen-Psaltis spacetime,
Proc. Workshop on Black Holes and Neutron Stars (2017).

\bibitem{Konoplya:2021slg}
R.~A.~Konoplya and A.~Zhidenko,
Shadows of parametrized axially symmetric black holes allowing for separation of variables,
Phys. Rev. D \textbf{103}, no.10, 104033 (2021).


\end{thebibliography}
\end{document}